\documentclass[12pt,preprint]{aastex}

\shorttitle{Spectral Components of SS~433}
\shortauthors{Gies et al.}

\begin{document}

\received{2001 July 18}
\accepted{}
 
\title{The Spectral Components of SS~433}

\author{D. R. Gies\altaffilmark{1,2}, 
        M. V. McSwain\altaffilmark{1,2},
        R. L. Riddle\altaffilmark{1,2,3},
	Z. Wang\altaffilmark{1,4},
        P. J. Wiita\altaffilmark{5}, 
        D. W. Wingert\altaffilmark{1}}

\affil{Center for High Angular Resolution Astronomy\\
Department of Physics and Astronomy \\
Georgia State University, Atlanta, GA  30303\\
Electronic mail: gies@chara.gsu.edu, 
mcswain@chara.gsu.edu, riddle@iastate.edu, wangzx@space.mit.edu,
wiita@chara.gsu.edu, wingert@chara.gsu.edu}

\altaffiltext{1}{Visiting Astronomer, Kitt Peak National Observatory,
National Optical Astronomy Observatories, operated by the Association
of Universities for Research in Astronomy, Inc., under contract with
the National Science Foundation.}

\altaffiltext{2}{Visiting Astronomer, University of Texas McDonald Observatory.}

\altaffiltext{3}{Current Address: Department of Physics and Astronomy,
Iowa State University, Ames, IA 50011}

\altaffiltext{4}{Current address: Center for Space Research,
Massachusetts Institute of Technology,
70 Vassar Street, Building 37, Cambridge, MA 02139}

\altaffiltext{5}{On leave at Department of Astrophysical Sciences, Princeton University}

\slugcomment{Submitted to ApJ}
\paperid{}


\begin{abstract}
We present results from new optical and UV spectroscopy of the 
unusual binary system SS~433, and we discuss the relationship of 
the particular spectral components we observe to the properties of 
the binary.   These spectral components include: \\
(1) The continuum spectrum which we associate with flux from 
the super-Eddington accretion disk and the dense part of its wind.
A FUV spectrum from HST/STIS made during the edge-on orientation 
of the disk places an upper limit on the temperature of an
equivalent blackbody source ($T< 21,000$~K for $A_V=7.8$) 
when combined with NUV and optical fluxes.  The continuum source has 
a radius of approximately half the binary separation 
which may be larger than the Roche radius of the compact star.   \\ 
(2) H$\alpha$ moving components which are formed far from the 
binary orbital plane in the relativistic jets.  
We confirm that these emission features appear as ``bullets''
at a fixed wavelength and may last for a few days.   We present 
a contemporary radial velocity curve for the precessional motion 
of the jets which includes the nodding motion caused by tidal 
interaction with the optical star.  \\
(3) H$\alpha$ and \ion{He}{1} ``stationary'' emission lines which 
we suggest are formed in the disk wind in a volume larger than 
the dimensions of the binary.   These lines vary on all time scales
and sometimes appear as P~Cygni lines.   We suggest that their 
radial velocity curves (which show greatest redshift at 
inferior conjunction of the optical star) result from 
an evacuation of the disk wind surrounding the optical star 
(caused by physical blockage, heating, or colliding winds). 
We argue that the wake of this interaction region causes an 
extended eclipse of the X-ray source (as seen in RXTE/ASM light curves).  \\
(4) A weak ``stationary'' emission feature we identify as a
\ion{C}{2} $\lambda\lambda 7231,7236$ blend that attains maximum 
radial velocity at the orbital quadrature of disk recession 
(like the velocity curve of \ion{He}{2} $\lambda 4686$). 
This is probably formed in outflow from the central region of 
the disk near the compact star. \\ 
(5) Absorption and emission features from outflowing clumps in the disk wind 
(seen most clearly in an episode of blue-shifted  \ion{Na}{1} emission).  \\
(6) We found no clear evidence of the absorption line spectrum of 
the optical star, although we point out the presence of 
\ion{He}{1} absorption features (blended with the stationary 
emission) with the expected radial velocity trend at the orbital 
and precessional phases when the star might best be seen.  \\
(7) A rich interstellar absorption spectrum of diffuse 
interstellar bands.  \\
The results suggest that the binary is embedded in an 
expanding thick disk (detected in recent radio observations) 
which is fed by the wind from the super-Eddington accretion disk.
\end{abstract}



\section{Introduction}                              

The unusual binary system, SS~433, is both of one of the most famous
and still most mysterious of the X-ray binary systems 
\citep{vdh81,mar84,cla85,zwi89}.
The basic scenario that has emerged from 24 years of investigation
is of an evolved binary caught in the act of extensive mass transfer.
The mass donor feeds an enlarged accretion disk surrounding a
neutron star or black hole companion, and some of the mass inflow
is redirected through the influence of the disk into oppositely
directed relativistic jets that are observed in optical and X-ray
emission lines and in high resolution radio maps.  There are two basic
clocks that dominate the spectral appearance and system dynamics.
The first is a 162~d periodicity that is observed in the extreme
radial velocities of the ``moving'' set of emission lines.  These red
and blue, mirror-symmetry motions are well described by the ``kinematical
model'' of emission from jets driven by a precessing disk 
\citep{fab79,mil79a,abe79}.

The second clock is the 13~d orbital period itself which was first
discovered through the radial velocity variations of a set
of ``stationary'' emission lines that presumably originate
in or near the disk \citep{cra80,cra81}.
The  H Balmer and \ion{He}{1} lines reach maximum velocity 
at superior conjunction of the X-ray source, and so this emission 
could originate in a gas stream from donor to compact star.
However, the high excitation emission line \ion{He}{2} $\lambda 4686$ 
has a different radial velocity curve with a maximum occurring 
at the quadrature when the compact object is receding 
\citep{cra81,kop89,dod91,fab97a}.   
This suggests that the \ion{He}{2} $\lambda 4686$
radial velocity curve traces the orbital motion of hot gas 
near the compact star.   The orbital mass function indicates that the
donor star mass is in excess of $8~M_\odot$ \citep{fab97a}, so we expect that
the surviving star is a massive OB object (although the clear spectral
signature of this star has eluded detection).   

The other major periodicity found in the radial velocities of the jets, 6.28 days,
corresponds to the time between the star's passage through the nodal line
of the disk, and this results from recurrent tidal deformations or
``nodding'' of the disk.  Thus, the combination of the precessional
and orbital periods has provided a very successful description of
the jet motions \citep{kat82,col94} and the photometric variations \citep{gor98a}.
Nevertheless, there remain a number of outstanding puzzles: 
the actual masses and nature of the donor and collapsed stars \citep{fuk98}, 
the jet formation process \citep{pan99,oku00,ino01}, the origin
of the disk precession \citep{wij99,ogi01}, and the evolutionary state 
of the binary \citep{kin00}.

The optical continuum spectrum of SS~433 has the shape 
of a highly reddened, Rayleigh-Jeans distribution 
implying an origin in a hot object \citep{mur80,che82,wag86,dol97}. 
The optical flux varies on the same precessional 
cycle as the jets \citep{gor98a} and results from the 
changing orientation of the precessing disk \citep{lei84,ant87,fuk98}. 
The large amplitude of the precessional variation 
indicates that the disk is the dominant flux source in 
the optical band.    The system also displays an 
orbital light curve with two unequal eclipses \citep{gor98a}, 
and the primary eclipse occurs when the X-ray source 
is eclipsed \citep{ste87}.   The relative depths of the eclipses 
indicate that the disk source has a characteristic 
temperature twice that of the optical star \citep{che82,lei84,ant87}.
Models of the optical light curve favor a
mass ratio $q= M_X / M_O = 0.4$ to 1.2 \citep{ant87,fuk98} 
which suggests a massive black hole companion, 
while models of the X-ray light curve \citep{ant92} 
indicate a smaller mass ratio indicative of a neutron star companion. 
This dilemma could be solved if the spectrum of the 
companion could be identified and its orbital velocity 
curve measured to provide a direct determination 
of the mass ratio. 

The other important observational challenge is 
to explore the faint ultraviolet part of the spectrum so as 
to better determine the temperature of the hot, super-Eddington 
disk and the nature of its UV radiation field 
which could play an important role in the mass outflow 
\citep{mil79b,pan99}.   \citet{dol97} succeeded in observing 
SS~433 in the near-UV with the High Speed Photometer aboard 
the {\it Hubble Space Telescope}.
Their work shows that the continuum polarization rises 
steeply towards the UV.  They sought but found no evidence of rapid 
flux variability in the UV that might be associated with the spin 
period of the collapsed star.   Unfortunately, their flux measurements 
at 2770~\AA ~are still far removed from the Wien peak of the 
expected flux distribution, and large uncertainties remain 
about the temperature associated with the continuum light. 
For example, \citet{dol97} fitted the optical and near-UV 
fluxes with a blackbody spectrum with $T=$ 72,000~K and 
$A_V=8.4$ while \citet{wag86} fitted optical spectrophotometry 
with $T=$ 45,000~K and $A_V=7.8$ (in both cases for the 
spectrum observed near photometric maximum when the 
disk normal is closest to our line of sight and the disk face
attains its largest projected area on the sky). 

Here we describe our recent program of spectroscopy of SS~433 
designed to answer some of these outstanding issues. 
We first discuss a program of extensive, moderate 
resolution, optical spectroscopy obtained with the 
Kitt Peak National Observatory Coude Feed Telescope and 
the University of Texas McDonald Observatory 2.1-m 
Telescope between 1998 - 1999 (\S2).   
The properties of the jet lines are described in \S3,
and the stationary lines are discussed in \S4.  
We examine in \S5 the continuum spectrum in the vicinity 
of H$\alpha$ to search for evidence of the 
optical companion.  We describe the first observation of the far-UV 
spectrum of SS~433 with the {\it Hubble Space Telescope} 
Space Telescope Imaging Spectrograph in \S6.  
Finally, we summarize in \S7 the
observed spectral components of SS~433 and their 
interpretation in framework of the binary model,  
along the same lines as done in earlier work by \citet{mur80} 
and \citet{che82}. 


\section{Optical Spectroscopy}                      

The optical spectra were obtained mainly with the Kitt Peak National 
Observatory 0.9-m Coude Feed Telescope between 1998 August and 
1999 November.   A summary of the different observing runs is 
given Table~1 which lists the beginning and ending heliocentric
dates of observation, the wavelength range recorded, the 
spectral resolution ($\lambda/\delta\lambda$), the number of 
spectra, and the instrumental configuration (codes for the 
observatory, telescope, grating, and detector). 
In the most common configuration with the Coude Feed, we used 
the short collimator, grating RC181 (in first order with a GG495 
filter to block higher orders), and camera 5 with a Ford $3072
\times 1024$ CCD (F3KB) as the detector.  We set the grating 
at two tilts to cover fully the wavelength range of the H$\alpha$ 
jet features in SS~433 (5900 -- 7750 \AA).   There were several  
departures from this standard arrangement.  The detector 
on the first night of the 1999 observations was a Texas Instruments 
$800 \times 800$ CCD (TI5), and we only recorded the spectrum 
with one grating setting in the immediate vicinity of the rest 
H$\alpha$ line.  The somewhat higher dispersion spectra from 1998 were 
made using the long collimator, grating B (in second order with order 
sorting filter OG550), camera 5, and the F3KB CCD.  Only one grating
setting was used, and thus the wavelength range is limited.  
Finally, we also obtained several high dispersion echelle spectra of 
SS~433 using the University of Texas McDonald Observatory 
2.1-m telescope and Sandiford Cassegrain Echelle Spectrograph (McCarthy et al.\ 1993). 
The detector was a Reticon $1200\times 400$ CCD (RA2) 
with $27 \mu$m square pixels which recorded 27 echelle orders 
covering the region blueward from H$\alpha$.  

\placetable{tab1}      

We usually obtained two consecutive exposures of 30 minutes duration 
and co-added these spectra to improve the S/N ratio.   We also observed 
with each configuration the rapidly rotating A-type star, $\zeta$~Aql,
which we used for removal of atmospheric water vapor and O$_2$ bands. 
Each set of observations was accompanied by numerous bias, flat field, 
and Th~Ar comparison lamp calibration frames. 

The spectra were extracted and calibrated
using standard routines in IRAF\footnote{IRAF is distributed by the
National Optical Astronomy Observatories, which is operated by
the Association of Universities for Research in Astronomy, Inc.,
under cooperative agreement with the National Science Foundation.}.
The Coude Feed spectra were traced, extracted 
(after subtraction of the sky background), and wavelength 
calibrated using the IRAF routine {\it doslit}, and the same 
tasks were done using {\it doecslit} for the McDonald Observatory
Sandiford echelle spectra.   The blaze function response of the
echelle spectra was removed by division of fits to spectra of 
hot single stars (or in some cases by fits of the flat field spectra). 
All the spectra were rectified to a unit continuum by the fitting 
of line-free regions using the IRAF task {\it continuum} 
(and in the case of the echelle spectra, the resulting orders were 
then linked together using the task {\it scombine}).
Note that the rectification process arbitrarily removes the continuum flux variations 
that are known to exist in the spectrum of SS~433, and the intensities
of the emission lines must be interpreted bearing in mind this 
renormalization.   The removal of atmospheric lines was done by 
creating a library of $\zeta$~Aql spectra from each run, removing 
the broad stellar features from these, and then dividing each target
spectrum by the modified atmospheric spectrum that most closely 
matched the target spectrum in a selected region dominated by 
atmospheric absorptions.  We also removed the most obvious 
air glow emission lines at this stage.  
The spectra from each run were then 
transformed to a common heliocentric wavelength grid. 
Finally, we transformed the spectra from all the runs onto 
a standard $\log \lambda$ wavelength grid between 5400 and 
7793\AA ~with a grid spacing equivalent to 20 km~s$^{-1}$ 
(this binning operation effectively lowered the resolution 
but increased the S/N of the echelle spectra). 


\section{Moving Emission Lines}                     

We begin by describing the appearance in our spectra of the 
``moving'' components of H$\alpha$ which originate in the 
relativistic jets of SS~433 \citep{ver93a,pan97}.
Our primary motivation is twofold: 
to establish an accurate velocity curve for the precessional 
cycle during our observations and to identify those times 
when the moving components cross and blend with the ``stationary'' lines. 
Figure~1 shows the spectral regions that included the moving lines
during our final run when the moving components were both 
strong and unblended with the stationary lines.   The left panel 
represents the nightly progression in the components from the 
approaching jet (H$\alpha -$) while the right panel 
shows the receding jet components (H$\alpha +$).  
Our results confirm the conclusions of \citet{bor87}, \citet{ver93a}, 
and others that these jet components appear as distinct 
``bullets'' of emission that appear suddenly at specific wavelengths 
and then decline on a time scale of a few days at the same position. 
There is often (but not always) a correspondence between the 
Doppler shifts and strengths of components from the approaching 
and receding jets.   We also confirm the dramatic variations 
in the intensities of the moving line components \citep{kop86,kop87};
for example, there were four nights near HJD~2,451,426 when 
any jet components were weaker than our detection threshold
(intensity $<6\%$ of the continuum). 
We inspected the individual spectra for evidence of moving
components of \ion{He}{1} $\lambda\lambda 5876, 6678, 7065$ 
(which have significant stationary lines), and weak but plausible 
emission components were only found in cases where the H$\alpha$ 
moving emission components were relatively strong. 

\placefigure{fig1}     

Following the approach of \citet{ver93a}, we found 
that the only practical method of measuring the Doppler shifts of 
these components was to make Gaussian fits of each clearly 
evident emission peak (using the IRAF routine {\it splot}). 
There are several cases where the emission profiles significantly 
depart from a Gaussian shape (probably due to blending 
of several components), and we used multiple Gaussian 
fits in those cases where an obvious intensity minimum 
separated neighboring peaks.   Our measurements of 
Doppler shift, $z$, equivalent width, $W_\lambda$, and 
profile full width at half maximum, FWHM, are listed in 
Table~2.  The measured $z$ values are plotted against time
in Figure~2.

\placetable{tab2}      
\notetoeditor{Table 2 to appear in full only in electronic form.}

\placefigure{fig2}     

The radial velocity curve displays the well-known 162~d 
precessional variation plus shorter term ``nodding'' motions 
that were first discussed by \citet{kat82}.  
\citet{kat82} show that a disk inclined to the orbital plane
will experience gravitational torques caused by the orbiting 
star that will vary with the nutational period, 
$(2 P_{\rm orb}^{-1} + P_{\rm prec}^{-1})^{-1}$, 
where $P_{\rm orb}$ and $P_{\rm prec}$ are the orbital 
and precessional periods.   This 6.28~d nodding motion 
dominates in the measurements from individual observing runs.   
We decided to fit our $z$ measurements using the 
\citet{kat82} model as a simple means to parameterize the 
complicated velocity curves.  Although more sophisticated 
models are available \citep{col94}, their use
was not warranted given the limited data available 
and the basic uncertainties surrounding the origin of the 
precessional motion \citep{wij99}.
 
We decided to constrain the fit at the outset with three 
well-determined parameters set according to the results of
\citet{mar89} based upon observations spanning a
decade: jet velocity, $v/c = 0.2602$, precessional angle 
between the jet and orbital plane normal, $\theta _0 = 19\fdg85$,
and inclination of precessional axis, $i = 78\fdg83$.
We took the precessional period, 
$P_{\rm prec} = 162.15$~d,  and the orbital period, 
$P_{\rm orb} = 13.08211$~d, from the extensive photometric 
light curve results of \citet{gor98a}.
We then made a non-linear least squares solution for the 
three remaining parameters in the \citet{kat82} model:
$t_{\rm or}$, the epoch of a time of binary quadrature, 
$\xi_0$, the angle of the disk line of nodes to the line of 
sight at time $t_{\rm or}$, and 
$P_0 = 2 \pi / \Omega_0$, a parameter that sets the 
semiamplitude of nodding perturbation motions in 
$\theta$ and $\xi$ (the azimuthal angle).  We applied weights to each measurement 
for fitting purposes of $1/(\triangle z)^2$ where we
estimated a Doppler shift error of 
$\triangle \lambda ({\rm \AA }) \approx 0.05 + 7 / |W_\lambda |$ 
from measurements in consecutive spectra.
This effectively weights the solution in favor of the
strongest emission components present. 
We arbitrarily selected a time close to the radial 
velocity extrema of the jet components as a starting value 
for $t_{\rm or}$.  Our fitting results are summarized in Table~3 
where we list the number of jet measurements used ($N$), epoch 
($t_{\rm or}$) for the angle to the line of nodes ($\xi_0$), 
the semiamplitude of the nodding motion 
in the azimuthal direction ($\triangle \xi$; the semiamplitude in 
the $\theta$ direction is $0.36 \triangle \xi$), and the 
root-mean-square deviation in $z$ of the fit 
(which is much larger than the measurement errors or indeed the 
widths of the emission components; see Fig.~6 below).  
The results from the H$\alpha -$ and H$\alpha +$ measurements 
are negligibly different, and we adopted the average values 
for a global solution which is plotted in Figure~2. 
The errors for this adopted solution were estimated as the 
absolute value of the difference between the results for the 
H$\alpha -$ and H$\alpha +$ sets. 

\placetable{tab3}      

The model fit is generally satisfactory although clear deviations
are present in some emission bullets (see also Fig.~4 in \citet{ver93a}).
According to the model fit, the time of the greatest velocity 
separation in the precessional period occurred on 
HJD~2,451,458.12 $\pm 0.23$ which is in good agreement with 
the precessional ephemeris of \citet{gor98a}  
(HJD~2,450,000.0 $+ 162\fd15 E$).  We will therefore
calculate precessional phase, $\Psi$, in the remainder of this paper
according to the ephemeris, HJD~2,451,458.12 $+ 162.15 E$, 
so that $\Psi = 0$ corresponds to greatest radial velocity separation 
(and greatest disk opening angle). 
Note that in earlier work by \citet{mar84} and others the precessional 
cycle was measured from the first radial velocity cross-over point
which occurs at $\Psi = 0.342$ in our ephemeris. 

The semiamplitude of the nodding motion 
is $\triangle z \approx 0.009$ which is in reasonable 
agreement with the ratio found by other investigators 
(0.007, \citet{kat82}; 0.010, \citet{kop87}; 0.006, \citet{gor98a}).
The time $t_{\rm or}$ in the \citet{kat82} model corresponds 
to a quadrature phase in the binary orbit ($\phi = 0.25$ or 0.75
where orbital phase $\phi$ is reckoned from the time of mid-eclipse). 
It is interesting to note that our fitted value of $t_{\rm or}$
occurs at $\phi = 0.83$ according to the light curve ephemeris of 
\citet{gor98a}, HJD~2,450,023.62 $+ 13.08211 E$ (which we use below
for orbital phase calculations) or 1.0~d later than expected.  
The time $t_{\rm or}$ is also 0.7~d later than the predicted time of 
maximum light associated with the nodding motions according to the 
nutational ephemeris of \citet{gor98a} (HJD~2,450,000.94 $+ 6.2877 E$).
These facts suggest that there is time lag in the tidal response 
between the outer and inner parts of the precessing disk \citep{gor98a}.

Table~2 also gives our measured equivalent width values for the 
H$\alpha$ bullets which are comparable in range to those found in 
previous results \citep{ver93a}.  \citet{pan97} 
describe evidence for systematic variations in bullet strength with 
precessional phase which they interpret as an emission anisotropy 
(stronger in the leading direction of the jets).  Unfortunately, 
we lack the photometry needed to transform from continuum based 
emission strength to absolute emission flux, so our results are not
suitable to address this issue.   However, we can investigate how the
equivalent width ratio 
$W_\lambda ({\rm H}\alpha -) / W_\lambda ({\rm H}\alpha +)$ 
varies with precessional phase \citep{asa86,kop87,ver93a,pan97}.  
We assume that the local continuum distribution 
varies as the Rayleigh-Jeans part of a hot 
blackbody spectrum, so that the physical emission flux ratio 
is approximately equal to 
$[W_\lambda ({\rm H}\alpha -) / W_\lambda ({\rm H}\alpha +)] ({\lambda +}/{\lambda -})^4$ 
where $\lambda -$ and $\lambda +$ are the observed bullet 
wavelengths \citep{ver93a}.  Figure~3 illustrates this 
physical ratio versus precessional phase (see also Fig.~6 of 
\citet{pan97}).  The solid circles represent 
ratios derived from the average for each run of all measurements 
with $|W_\lambda | >10$~\AA .   
Because the jets have attained modestly relativistic speeds, radiation from 
the approaching jet should appear boosted compared to the receding jet, 
and the ratio is expected to vary with precession phase as 
$[(1~+~z+ ) / (1~+~z- )]^4$ ({\it solid line}) if the bullets 
can be considered as discrete components \citep{urr95}.  
The observed increase in the ratio when the jets are closest to our 
line of sight (at $\Psi = 0$) was also found by \citet{asa86} and \citet{pan97}, 
and appears to be close to the predicted amount of boosting. 
On the other hand, \citet{pan97} argue that the bullets 
in SS~433 should be regarded as a stream of individual gas clouds, 
in which case the the exponent in the above relation is decreased 
by one power of the Doppler factor because of Lorentz contraction
\citep{urr95}.   If so, then relativistic boosting ({\it dotted line})
is insufficient to account for observed overluminosity of 
the approaching bullets, and the 
relative weakness of the receding bullets might be due to some
increase in intervening opacity (perhaps from the disk) near 
$\Psi = 0$ \citep{pan97}.  Note that these expressions are based upon 
the assumption that the lines are broad enough that transformations 
appropriate for broad-band, rather than fixed frequency, emission 
are being considered. 

\placefigure{fig3}     


\section{Stationary Emission Lines}                 

The most striking features in our optical spectra are 
the so-called ``stationary lines'' of H$\alpha$ and
\ion{He}{1} $\lambda\lambda 5876, 6678, 7065$ 
\citep{cra81,mar84,fal87,zwi89,kop89,ver93a,fab97a}. 
These lines do, in fact, display small radial velocity shifts 
that are related to orbital phase, but the interpretation 
of the variations is still a matter of debate (see below). 
Here we describe the overall appearance of these 
lines and their variations both between and within runs. 

The average H$\alpha$ profiles for each run are plotted 
as a function of radial velocity and date in Figure~4. 
The profile is characterized by a central emission core 
and broad wings that extend to $\pm 2000$ km~s$^{-1}$. 
Both core and wings are time variable, and during one 
run (around HJD~2,451,426) the feature developed into a conspicuous 
P~Cygni profile with a blue-shifted absorption trough. 
Other investigators \citep{cra81,kop89,fab97a} 
have found similar P~Cygni profiles
usually during precessional phases when the jet axis is 
normal to our line of sight (and the disk 
is nearly edge-on) which occurs in the range $\Psi \approx 0.3 - 0.7$,
but the P~Cygni feature appears in our data after this interval. 
We also show in Figure~4 the mean profile from our 1998 run 
inserted at a date corresponding to its precessional 
phase.  There is no clear evidence in our set of spectra 
of strength or shape variations that are related to 
precessional phase.   

\placefigure{fig4}     

The mean profiles of \ion{He}{1} $\lambda\lambda 5876, 6678, 7065$ 
are illustrated in a similar format in Figure~5.   They 
generally have a similar appearance to one another and to the 
H$\alpha$ core, and the P~Cygni episode noted above in 
H$\alpha$ is also clearly evident in all three \ion{He}{1} lines
as well as \ion{O}{1} $\lambda 7772$ 
(which sometimes has absorption as deep as 50\% of the continuum). 
The P~Cygni shape is also seen in two earlier runs in 
\ion{He}{1} $\lambda 5876$ alone.   
The \ion{He}{1} profiles sometimes show a double-peaked 
emission maximum which also occurs in H$\alpha$.  
The broad emission wings seen in H$\alpha$ are generally 
not found in the \ion{He}{1} profiles, except during 
the final run when the wings were very strong in H$\alpha$.
We also note the development during the last two runs 
of blue-shifted \ion{Na}{1} $\lambda\lambda 5890, 5896$ 
emission which appeared at a radial 
velocity of $-27$ km~s$^{-1}$ (compared to the mean 
interstellar Na~D absorption velocity of $+35$ km~s$^{-1}$).

\placefigure{fig5}     

The evolution of the H$\alpha$ profiles on shorter time scales
is shown for two runs in Figures 6 and 7.   Both plots 
indicate that the profile is constantly varying with somewhat 
more rapid (night-to-night) changes occurring in the broad wings.  
Two examples of the wing variations are particularly 
noteworthy.  Figure~6 shows that the blue emission wing 
was very strong around HJD~2,451,357, and several, weak 
absorption troughs are visible.  One of these troughs 
apparently migrated bluewards during this run at a 
nearly constant acceleration of $\approx -30$ km~s$^{-1}$~d$^{-1}$ 
({\it dotted line}), reaching $-539$ km~s$^{-1}$ in the 
last spectrum of the run.  Figure~7 shows another outwardly 
accelerating feature, this time seen in emission in the 
red wing of the profile ({\it dotted line}).   This 
feature was seen to grow and decline over a period of 3~d
with an acceleration of $\approx +92$ km~s$^{-1}$~d$^{-1}$, 
and it was last clearly visible at a radial velocity of 
$+1234$ km~s$^{-1}$.   These kinds of outwardly moving 
sub-features are also observed in the emission lines 
of Wolf-Rayet stars \citep{lep00} where they are 
interpreted as evidence of clumping in the wind outflow. 

\placefigure{fig6}     

\placefigure{fig7}     

We measured the equivalent width ($W_\lambda$), FWHM, 
and radial velocity for the stationary lines of 
H$\alpha$ and \ion{He}{1} $\lambda\lambda 5876, 6678, 7065$, 
and the results are collected in Tables~4 and 5, respectively.  
The profiles are complex and asymmetric, and so we made two
radial velocity measurements, an intensity weighted centroid
of the entire profile ($V_{\rm centroid}$) and a parabolic fit 
of the upper third of the emission peak ($V_{\rm peak}$). 
Several of the \ion{He}{1} lines 
posed special problems.  We measured only the emission portion of 
those with P~Cygni profiles, and we did not include the weak 
and broad emission component that appeared in the last run. 
The Na~D lines acted as an arbitrary boundary for the 
integration of equivalent width for \ion{He}{1} $\lambda 5876$.
We then searched for correlated variations in the measurements 
with the orbital and precessional phases, and except in the case
of $V_{\rm peak}$ (see below), we found no compelling evidence of 
variability related to either of these clocks.  The lack of correlation 
is surprising since other investigators have claimed orbital 
and precessional related velocity and strength variations in 
earlier observations \citep{zwi89,kop89,fab97a}.    There is no 
apparent change in H$\alpha$ emission equivalent width 
during the photometric primary eclipse, in agreement with 
the results of \citet{gor97}. 
There is also no evidence of precessional related variations 
in H$\alpha$ FWHM, which would be expected if the emission 
formed in a disk which changes orientation with precession.
For example, according to the kinematical model for the 
jets \citep{mar89}, we would expect the sine of the 
angle between the disk normal and observer to be about 
0.86 at greatest disk opening (precessional phase $\Psi =0$) 
compared to 1.00 at the disk edge-on orientation 
($\Psi = 0.34$ and 0.66), and therefore, lines broadened 
by disk rotation would vary in width concurrently. 
However, even if such variations are present, they are overwhelmed by larger, 
temporal variations in width (for example, an increase by a factor of 2 
over the duration of the final run). 

\placetable{tab4}      
\notetoeditor{Table 4 to appear in full only in electronic form.}

\placetable{tab5}      
\notetoeditor{Table 5 to appear in full only in electronic form.}

The one correlation that did emerge is the variation in $V_{\rm peak}$ 
with orbital phase.  We found that in each run this 
radial velocity attained a different maximum near orbital phase
$\phi =0$, the time of primary eclipse.  In order to compare the results
from different runs, we applied an offset to each set 
of radial velocities which was calculated to minimize the scatter 
around a common sinusoidal curve.   The sinusoidal fit 
parameters are given in Table~6 where $K$ represents the 
semiamplitude of variation, $\phi$(max.) is the orbital phase of 
maximum radial velocity, $V_n$ is the systemic radial velocity 
found in run number $n$ (see Table 1), and r.m.s.\ is the 
root mean square of the residuals from the fit.  The differential 
velocity measurements and fits are illustrated in Figure~8. 
Our results agree reasonably well with earlier estimates ($K=73 \pm 4$ km~s$^{-1}$,
\citet{cra80}), but there appears to be considerable cycle-to-cycle 
variations in semiamplitude which may explain differences found 
between investigators.   As noted first by \citet{cra81}, the
occurrence of a velocity maximum at orbital conjunction implies that
we are {\it not} measuring Keplerian motion in these features. 
\citet{fab97a} suggests that the systemic velocity of the stationary 
lines varies in step with precessional phase (see his Fig.~2), 
and our results confirm that the systemic velocities tend to be lower 
near $\Psi = 0$.   This may reflect differing amounts of subtle,  
blue-shifted absorption is the wind outflow that has greatest effect 
when the disk has an edge-on orientation ($\Psi = 0.3 - 0.7$). 

\placefigure{fig8}     

\placetable{tab6}      

\citet{fab97a} and others have demonstrated that the high excitation 
\ion{He}{2} $\lambda 4686$ line attains maximum radial velocity 
at orbital phase $\phi = 0.75$ which is consistent with the expected 
orbital motion of emitting gas near the X-ray source.   We found only 
one example of such Doppler shifts in the spectral range we observed, in 
the weak emission line blend of \ion{C}{2} $\lambda\lambda 7231, 7236$ 
(see Fig. 10 in the next section).   This feature is seen in emission 
in stars with strong winds, such as cool W-R stars and luminous 
blue variables like P~Cygni \citep{sta93}.   The line is difficult to 
measure because it falls in a region of strong atmospheric lines
and lies close to the weak \ion{Ne}{1} $\lambda 7245$ emission line
from city lights.  Nevertheless, it appears double-peaked in our 
best spectra with the same separation as the \ion{C}{2} features, which 
strengthens the case for its identification.  We measured centroid 
velocities for this feature in those spectra where the emission 
was readily detected, and these radial velocities are plotted 
versus orbital phase in Figure~9 (for an assumed rest wavelength 
equal to the mean of the two, 7233.9~\AA ).   Unlike the H$\alpha$ 
and \ion{He}{1} stationary line velocity curves, this line appears
to reach maximum velocity at phase 0.75 (although the actual maximum 
velocity may vary from cycle to cycle).   A sinusoidal fit with the 
maximum set at this value yields a semiamplitude, $K = 162 \pm 29$ 
km~s$^{-1}$ and a systemic velocity, $V_0 = 200 \pm 20$ km~s$^{-1}$, 
in reasonable agreement with the velocity curve of 
\ion{He}{2} $\lambda 4686$ \citep{fab97a}.   If we assume 
that the velocity curve represents orbital motion, 
then this feature, like \ion{He}{2} $\lambda 4686$,  
probably forms close to the X-ray source in a region surrounding 
the disk center.    

\placefigure{fig9}     

Our results for the strong H$\alpha$ 
and \ion{He}{1} stationary lines are best explained if these
lines originate in a wind outflow from the 
accretion disk \citep{gor97} rather than in the disk itself 
or in a gas stream.   The overall H$\alpha$ shape 
varies from a strong, rounded top emission line to a weaker, 
P~Cygni profile which certainly implicates a 
mass outflow.  \citet{cas70} showed in some of the first models for 
winds from stars that optically thick outflows produce strong, rounded-top 
profiles with no blue absorption while P~Cygni profiles are 
associated with more optically thin winds.    Modern models for 
winds from the accretion disks in cataclysmic variables
\citep{kni95,fel99a,fel99b} show that winds from disks can also produce 
P~Cygni and single peaked emission lines depending in part on 
disk orientation.   The change between P~Cygni and single peak 
appearance in our data does not appear to be strictly related to 
precessional phase (disk orientation), so the variations we 
observe in the stationary lines of SS~433 probably result 
at least partially from temporal changes in the wind structure
instead of changes in orientation. 

Indeed the observations point to a wind outflow that is constantly 
changing.  We observe long term changes (weeks) in overall strength 
and radial velocity that probably correspond to global changes in the 
disk-wind outflow, plus we find rapid variations (days), such as the 
accelerating features in the H$\alpha$ wings, that suggest the 
development of clumping or other shock structures in the wind. 
The appearance of blue-shifted Na~D emission in the last runs 
suggests the presence of outflowing structures forming at 
a large distance from the binary where the gas can cool. 

The variation in $V_{\rm peak}$ with orbital phase (Fig.~8) is probably 
due to the influence of the mass donor star on the structure of the 
accretion disk's wind \citep{gor98a}.  The companion is probably 
a large and hot star \citep{gor98a,kin00} with significant 
radiative flux and possibly a wind of its own.   This star
could effectively remove the disk-wind outflow in its vicinity 
by direct blockage of the flow \citep{gor98a}, 
through the formation of a colliding winds bow shock between 
the stars \citep{che95}, or through photoionization \citep{hat77} 
(see the case of Cyg X-3; \citet{van96}).  
In the case of H$\alpha$, the reduction in flux could be caused 
by heating of the gas in the vicinity of the donor star and 
a subsequent decrease in H$\alpha$ emissivity \citep{ric98}. 
Thus, when we observe the system near primary eclipse, $\phi =0$, 
a portion of disk-wind volume in the foreground surrounding the 
companion is absent, and there is a net decrease in blue-shifted 
emission (yielding a red-shift in the remaining emission envelope). 
Likewise, at $\phi =0.5$, the evacuated region occurs in the 
region moving away from us, so that the red-shifted emission declines. 
The fact that this $V_{\rm peak}$ variation appears 
different from orbit to orbit suggests that there are temporal 
variations in the boundary or shape of the evacuated region.  
Note that the gas stream model for the H$\alpha$ stationary line 
\citep{cra81,kop89} is not viable, since a Roche lobe overflow stream 
from the donor would be eclipsed near or shortly after $\phi =0$
which we do not observe.   

Disk-wind models for cataclysmic variables \citep{kni95,fel99a,fel99b} 
show that the inner disk produces a faster flow, more 
collimated with the disk normal, while the outer portions 
produce a slower, dense wind that flows closer to the disk plane. 
We speculate that the wings and core of the stationary 
H$\alpha$ emission correspond to these two outflow regimes, 
respectively, since the wings have the large velocities and rapid 
variability expected in the inner disk, high speed wind.  
The outer disk wind may be the source of an extended circumbinary 
gas torus, and such a region was recently discovered by \citet{par99}
in radio emitting clouds normal to the jet axis. 


\section{The Continuum Spectrum}      

The spectral features described thus far are probably formed 
in outflows from the system, and it is important to check for 
any features that might form in the donor star or disk.  
Such features would display modest Doppler shifts 
(probably $<200$ km~s$^{-1}$), and so we formed a simple 
average spectrum to search for any weak features associated 
with the star or disk.   We first subtracted out the Gaussian fits 
made of the moving H$\alpha$ features (Table~2), and then 
at each wavelength point we formed the mean and standard 
deviation, $\sigma$, from all our available spectra, and 
finally we formed a second 
mean after deletion of any points more than $2\sigma$ away from 
the first mean.  We then made a careful inspection of the spectrum for 
interstellar absorption lines (mainly diffuse interstellar bands) 
from the lists of \citet{her75}, \citet{her91}, \citet{mor91}, and especially 
\citet{gal00}, and an interstellar spectrum was formed by extracting the 
mean SS~433 spectrum in the immediate vicinity of each interstellar 
absorption line.   This spectrum of interstellar features appears 
in Figure~10, and it bears a strong resemblance to the mean 
interstellar spectrum of heavily reddened stars published by 
\citet{jen94} (see their Fig.~4).  The mean SS~433 spectrum 
divided by the interstellar spectrum appears in Figure~10 along 
with a sample night sky absorption spectrum and the spectrum of 
a hot star, 9~Cep (B2~Ib) (with interstellar lines removed).  
All of these spectra were smoothed by convolution with a 
Gaussian filter of FWHM = 100 km~s$^{-1}$ (5 pixels).  

\placefigure{fig9}     

The red spectral range of SS~433 is marked by the strong 
``stationary'' features (H$\alpha$, 
\ion{He}{1} $\lambda\lambda 5876$, 6678, 7065) plus 
the weaker features of \ion{C}{2} $\lambda\lambda 7231,7236$,
\ion{He}{1} $\lambda 7281$, and \ion{O}{1} $\lambda 7772$.  
The remaining features may represent very weak stationary lines
(see for comparison the wind outflow lines in the P~Cygni 
spectral atlas by \citet{sta93}), but they could also be due to  
residual flat fielding problems, 
problems in dividing out the night sky absorption (near 6868 and 7605 \AA ), 
air glow emission lines (many weak lines such as OH at 7712 \AA ), 
incomplete removal of broad interstellar features (such as the 
broad profile near 6172 \AA ; \citet{gal00}), 
and incomplete removal of weak jet features. 
There are no stellar absorption lines present deeper than 
5\% of the continuum with the exception of the P~Cygni trough 
of \ion{O}{1} $\lambda 7772$.
The donor star is expected to contribute an increased fraction 
of the flux during primary eclipse, and so we also 
examined an eclipse spectrum made at 
HJD~2,451,357.8 but found no
evidence of additional absorption lines. 

The spectrum of 9~Cep (B2~Ib) illustrated in Figure~10 shows the kinds of spectral 
lines we might expect to find if the donor was a typical, early B-type 
star.   The strong lines in such a spectrum are the same set 
that forms the stationary emission lines in SS~433, 
and since the emission from outflows dominates in these features, 
they cannot be used easily to find the donor's spectrum. 
The only other features available in this region cluster 
near 5700 \AA ~(including \ion{N}{2} $\lambda\lambda 5666,5679,5710$,
\ion{Al}{3} $\lambda\lambda 5696, 5722$, and \ion{Si}{3} $\lambda 5739$), 
but there are no corresponding absorption lines in the SS~433 spectrum.  
If the donor was an O-type star, we might expect to find 
transitions of \ion{He}{2} $\lambda 5411$, \ion{O}{3} $\lambda 5592$, 
\ion{C}{3} $\lambda 5696$, and \ion{C}{4} $\lambda\lambda 5801,5812$
\citep{con74}, but there is no evidence of these lines in the 
SS~433 spectrum either.   A cooler donor might more easily escape 
detection in this spectral region, but even in the case of a cooler 
B-type star, we might expect to find transitions like 
\ion{Si}{2} + \ion{Mg}{2} $\lambda 6347$, \ion{Si}{2} $\lambda 6371$, and 
\ion{Ne}{1} $\lambda 6402$ \citep{gal00}.   Again we find no trace of these
absorption lines in the SS~433 spectrum. 

We also used a Doppler tomography algorithm \citep{bag94} to 
make trial reconstructions of the donor star spectrum for 
assumed values of semiamplitude, $K_O = 80 - 320$ km~s$^{-1}$ 
\citep{ant87}, $K_X = 175$ km~s$^{-1}$ \citep{fab97a}, and 
flux ratio, $F_O / (F_O + F_X) = 0.25$.  
We restricted the sample to the KPNO spectra from 1999
to insure a homogeneous set of spectra, and the jet components 
were divided out prior to reconstruction.   The algorithm assumes that
the spectrum features can be assigned to a disk or donor component 
with radial velocity variations defined by the photometric phase 
of observation and assumed semiamplitudes, and so the strong stationary 
lines, which follow neither curve, are arbitrarily divided between the
disk and star components in the reconstruction.   We searched for 
stellar lines away from these emission artifacts, but again there 
were no stellar absorptions visible with depths exceeding 10\% of the 
continuum.  Since O- and B-type stars generally have metallic 
line depths of this order in this red region (for moderate projected 
rotational velocities), our null detection of such absorption lines 
implies that a normal companion can contribute no more than 25\%
of the continuum flux in this part of the spectrum, in agreement with 
estimates from analyses of the optical light curve \citep{lei84,ant87,san93}. 
However, we suspect that the photosphere of the donor star 
remains hidden behind the veil of the disk wind, and so such limits
on the donor's flux contribution based on line depths should be 
treated with caution.  


\section{UV Spectrum}                               

Our original goal in this program was to obtain 
{\it Hubble Space Telescope (HST)} Space Telescope 
Imaging Spectrograph (STIS) observations of the 
far ultraviolet spectrum at several points throughout 
the precessional period.  However, our single observation 
(made on 1999 July 3) showed no easily detectable UV flux, 
and so subsequent planned observations were abandoned. 
Here we discuss this one null result and its implications 
for the UV spectral flux distribution. 

The STIS observation was made with the first order G140L grating 
which covers the range 1150 -- 1700 \AA ~with a spectral 
resolution of $\lambda / \Delta \lambda = 1000$ 
(average reciprocal dispersion of 0.6 \AA ~pixel$^{-1}$). 
The star was centered in the large $52\times 2$~arcsec aperture, 
and the spectrum was recorded on the FUV-MAMA detector in 
ACCUM (integration) mode.    
The spectrum was made in three sub-exposures for a 
total of 7534~s of targeted exposure.   
The spectra were reduced with the standard STIS pipeline 
software using contemporary background, flux, and 
wavelength calibrations \citep{voi97}.   
The final product we examined was 
a co-added and flux calibrated image in rectified 
spatial and wavelength coordinates.   

The image of the UV spectrum is dominated by broad bands of 
geocoronal emission (primarily near Ly$\alpha$ $\lambda 1216$ 
and \ion{O}{1} $\lambda\lambda 1302, 1356$) that span the 
full spatial dimension, and these are superimposed upon 
a very low level, diffuse background \citep{bro00}.  
A visual inspection of the wavelength-summed spectrum 
along the spatial dimension showed no evidence of 
an obvious peak near the expected position of the stellar 
spectrum, nor any sign of an emission peak near Ly$\alpha$. 
Nevertheless, we attempted to extract the 
spectrum by first fitting a second order polynomial to 
the background at each wavelength point over the spatial 
pixel range $(-100:-20,+20:+100)$ relative to the predicted 
stellar spectrum position, and then this fit was subtracted 
and the net flux was integrated over 11 spatial pixels centered
on the position of the stellar spectrum.   The spatially 
integrated flux was then converted to units of 
erg~cm$^{-2}$~s$^{-1}$~\AA $^{-1}$ using the averaged, 
point source, aperture throughput conversion factor
\citep{voi97}.  We then formed the average flux in 
wavelength bins of width approximately 50~\AA ~that 
were selected to include regions of comparable background 
noise (for example, by selecting bin boundaries that 
coincide with the ranges of the geocoronal emission 
features), and we also calculated the standard deviation 
of the mean, $\sigma_\mu$, in each of these samples (Table~7).  
In no case did we find a mean that exceeded $3\sigma_\mu$,
and so we cannot claim that the UV flux of SS~433 
was actually detected.  Instead, we set upper limits
for any such flux as the sum of the mean flux 
plus twice the standard deviation of the mean.  
Even if the stellar spectrum was misplaced from the
expected spatial position, the upper limits should 
reflect the noise character found in nearby parts of the image.  
These upper limits are plotted in Figure~11. 

\placefigure{fig10}     

\placetable{tab7}       

The only reliable detection of SS~433 in the UV band was 
made with the {\it HST} High Speed Photometer (HSP) by 
\citet{dol97}, and it is worthwhile comparing our 
far-UV limits with their near-UV measurements. 
\citet{dol97} observed SS~433 at epochs when the 
system is generally brighter and fainter in the 
precessional cycle (corresponding to times when the 
disk face is maximally opened in our direction and 
edge-on, respectively), and we show in Figure~11 
the average flux they detected for both states 
in a filter centered on 2770~\AA .   There exists only one 
low dispersion spectrum of SS~433 in this spectral region 
in the archive of the {\it International Ultraviolet 
Explorer Satellite (IUE)} (LWR4698, obtained by Dr.\ A.\ Underhill), 
and we also show in Figure~11 the mean flux measured in 
three bins of 50~\AA ~width in the best exposed part of 
that spectrum.   The optical spectrophotometry from
\citet{wag86} is illustrated for average bright 
($0.8<\Psi <0.2$) and faint ($0.2<\Psi <0.8$) states. 

\citet{dol97} fitted the bright state 
near-UV and optical fluxes with a reddened blackbody 
spectral distribution for $T=$ 72,000~K and $A_V = 8.4$, 
and we have used these parameters to calculate the 
predicted far-UV fluxes using the interstellar extinction 
curve defined by \citet{fit99} 
(for an assumed $R=A_V / E(B-V) = 3.1$).  
The extrapolation of the 
\citet{dol97} flux distribution into the far-UV 
({\it dotted line}) predicts fluxes that we would have 
detected around 1500~\AA .   Furthermore, the 2770~\AA 
~and red fluxes for the high state do not match the curve 
particularly well.  \citet{wag86} finds a better match 
with a lower extinction, $A_V = 7.8 \pm 0.5$, which 
agrees with estimates from other investigators \citep{mur80,che82}.   
His fit of the high state optical fluxes indicates 
a blackbody temperature of $T=$ 45,000~K, and this 
curve ({\it dashed line}) also predicts UV fluxes that we 
would have detected.   However, the STIS spectrum 
was obtained on HJD~2,451,362.99 which corresponds to 
precessional phase $\Psi = 0.41$, close to the disk 
edge-on configuration when the system is fainter. 
In fact, a comparison of the counts in the STIS acquisition 
image with those predicted from the STIS Exposure Time Calculator 
\footnote{http://garnet.stsci.edu/STIS/ETC/stis$\underbar{~}$acq$\underbar{~}$etc.html}
indicates that at the time of our observation 
SS~433 was only 1.2 times brighter than the 
average faint state spectrum of \citet{wag86}. 
Thus, a better comparison might be made with the 
faint state fit given by \citet{wag86} 
(for $T=$ 21,000~K and $A_V = 7.8$).
This fit ({\it solid line}) predicts fluxes which fall 
just below our STIS detection limits.   
A low temperature like this is 
more consistent with the UV flux observed in another 
X-ray binary, LMC~X-3 (fit with $T=$ 30,000~K), in which the 
disk out shines the stellar flux contributions \citep{cow94}. 
Note that the spectral flux distribution of a super-Eddington 
accretion disk may depart from blackbody expectations 
especially at shorter wavelengths \citep{lip99,oku00}. 


\section{Discussion}                               

Our objective in this final section is to 
relate the various spectral components discussed above to 
specific parts of the binary system.  The well established 
components of SS~433 include the mass donor star (which 
must be close to Roche-filling in order to 
transfer mass to the the disk and 
power the X-ray source), the relativistic star (neutron star 
or black hole), the super-Eddington accretion disk and its wind, and 
the jets.   The binary period is well known from the optical 
light curve \citep{gor98a}, and the system inclination 
is determined from the kinematical model for the jets \citep{mar89}.   
However, there still exists considerable uncertainty about the binary 
orbital velocity curves (and thus the size of the semimajor axis)
and mass ratio.   The radial velocity curve of the disk 
is measured indirectly through the Doppler shifts in the 
\ion{He}{2} $\lambda 4686$ emission line which is probably 
formed in a hot gas outflow near the base of the jets \citep{gor97,fab97a}.
The shape of the \ion{He}{2} $\lambda 4686$ velocity curve 
changes with precessional phase \citep{fab97b}, and 
\citet{fab97a} argues that the curve obtained near maximum 
disk opening angle ($\Psi = 0$) offers the best estimate 
of the disk's orbital motion.  The observed semiamplitude, 
$K \approx 175$ km~s$^{-1}$, yields a mass function for the
system inclination ($i=79^\circ$) of approximately
$M_O^3 / (M_O + M_X)^2 = M_O / (1 + q)^2 = 7.7 M_\odot$,
where $q=M_X/M_O$.   The semimajor axis is then 
given by $a = (1+q) a_X = (1+q)~ 3.2\times 10^{12}$~cm (or 
$(1+q) 46 R_\odot$).   Analysis of the optical light curve 
yields mass ratio estimates from $q=0.4$ to 1.2 \citep{ant87,fuk98}
while models of the X-ray light curve suggest a lower range, 
$q=0.15$ to 0.25 \citep{ant92}.   
 
The optical light curve \citep{gor98a} provides some guidance 
about the proportions of the continuum flux that originate 
in the disk and star.   The mean, out of eclipse, $V$-band 
magnitude exhibits a 0.60 mag variation through the 
precessional cycle (that tracks with the motion of the jets) 
which indicates that changes in disk orientation cause 
a modulation of 55\% of the total flux.  A variation this
large is only possible if the disk is the dominant source of 
light in the optical.   Furthermore, even during the center 
of the primary eclipse, the precessional light curve varies 
by 0.41 mag \citep{gor98a} which suggests that the disk 
contributes a considerable fraction of the flux even during
these partial eclipses.   We can estimate the size of the 
continuum forming region using the blackbody fits to the 
flux distribution given in \S6. 
The continuous spectrum fits shown in Figure~11 
are normalized for a projected disk radius  
of $1.4\times 10^{12}$~cm ($20~R_\odot$) and 
of $2.3\times 10^{12}$~cm ($33~R_\odot$) 
for the high and low states, respectively, 
based on the temperature fits of \citet{wag86} and 
an assumed distance of 4.85~kpc \citep{ver93b}. 
When the disk face attains its maximal opening towards us in the 
high state, we may be seeing the hotter, central regions 
of the disk, while during the edge-on, low state, we 
view more flux contributions from the outer (larger) 
regions of the disk.  
The low state radius is approximately half the semimajor axis 
for reasonable values of the mass ratio, and so we should 
admit the possibility that the continuum forming part of the 
disk may extend beyond the Roche lobe of the relativistic object.  

Next we turn to the formation sites of the emission lines. 
The moving emission lines form in the relativistic 
jets (\S3), and based upon their average lifetime and 
outflow velocity, these probably form at a distance of 
$\approx 4\times 10^{14}$~cm ($6000 R_\odot$) from the 
central engine, i.e., $\approx 100\times$ further out 
than the binary separation \citep{pan99}.  
On the other hand, we argued above (\S4) 
that the stationary emission lines form in the disk wind 
with dimensions more comparable to the system separation. 
\citet{mur80} show that the remarkable strength of the 
H$\alpha$ emission peak relative to the surrounding continuum 
implies that the H$\alpha$ emission is formed over a radius 
at least $10\times$ larger than the continuum source, 
which implies that line emitting regions of the disk wind 
extend far beyond the binary system.   The optical 
star must be orbiting within a dense part of this outflow, 
and it apparently sculpts out a wind evacuated region in 
its immediate vicinity (through physical blockage, 
ionization of the nearby gas, or colliding winds) 
that leads to the orbital radial velocity variation 
we observe (Fig.~8).  
 
Evidence of the interaction between the disk wind 
and star is probably also found in the X-ray light curve.   
We constructed a contemporary X-ray light curve 
using raw counts of the source from the 
Rossi X-ray Timer Explorer Satellite's
All Sky Monitor (RXTE/ASM) 
instrument\footnote{http://xte.mit.edu/} \citep{lev96}. 
The ASM is sensitive to X-rays in the 1.5 -- 12 keV range, 
and some 1673 measurements are available of SS~433 
over the time interval JD 2,450,089 -- 2,452,054. 
We selectively deleted those observations 
of high flux made during flaring events and 
those with net negative fluxes (null detections 
perhaps associated with X-ray dimming events 
comparable to those observed in the optical; 
\citet{gor98a}) to arrive at 1363 observations 
that correspond to the quiescent X-ray state. 
Figure~12 ({\it top panel}) illustrates the X-ray fluxes 
binned into 10 intervals of precessional phase (based 
on our precessional ephemeris).   The X-ray flux clearly 
reaches a maximum near $\Psi = 0$, the phase of maximum 
disk opening, as has been found in earlier work 
\citep{yua95}.  If we restrict the ASM measurements to 
$\Psi = 0.8 - 0.2$, the phase interval surrounding the maximum, 
we also see evidence of X-ray flux variations nearly in 
step with the optical variations on the nutational 
periodicity, 6.2877~d \citep{gor98a}, caused by the nodding 
motions of the disk (Fig.~12, {\it lower panel}). 
The X-ray variations on the orbital cycle are shown in 
the central panels of Figure~12 for the disk opening ($\Psi = 0.8 - 0.2$)
and disk edge-on ($\Psi = 0.2 - 0.8$) configurations during the precessional cycle. 
A significant X-ray eclipse is only observed in the disk opening state
({\it second panel from the top}), and the primary eclipse 
appears to be delayed or extended beyond the expected time 
around optical star inferior conjunction ($\phi = 0$),  
referred to as a ``second eclipse'' by \citet{yua95}. 
We suggest that this extended eclipse is caused by 
dense, X-ray opaque gas that trails behind the optical star 
as it plows through the disk wind.  Earlier interpretations 
of the X-ray light curve required a large optical star 
(and therefore small mass ratio, $q$) in order to explain the long 
duration of the eclipses \citep{ant92}, and it would be rewarding 
to revisit the fits of the light curve taking full account 
of a wake induced extension of the eclipse.  We note that a minor 
additional X-ray eclipse occurs at $\phi = 0.5$ when the disk 
is in the foreground.   This suggests that an additional 
source of X-rays exists between the stars, and we tentatively 
suggest that a small fraction of the X-ray flux forms in 
a colliding winds shock near the optical star \citep{che95}. 

\placefigure{fig11}     

All these observations suggest that the central binary 
is embedded in a large equatorial disk that is formed 
by the disk wind and extends far beyond the binary itself. 
Large outflowing disks are found in other massive 
binary systems, in particular, the W~Serpentis class 
\citep{tar00} which includes such noteworthy examples 
as $\beta$~Lyr \citep{lin00} and RY~Scuti \citep{smi99}. 
In the case of $\beta$~Lyr, the disk is so thick that it 
blocks the photospheric flux of the mass gainer.  
The outer portions of the disk in SS~433 have 
recently been detected in the radio continuum by 
\citet{par99} at a projected separation of $\approx 50$~mas 
or $4\times 10^{15}$~cm ($50,000~R_\odot$) away from the 
central binary.   Variations in the inner portion of the disk 
may be responsible for the irregular changes in the red 
flux component discussed by \citet{gor98b}. 

The disk probably widens significantly 
with distance from the binary due to the changing
orientation of the disk wind source with precession, 
and we suggest that the disk blocks our view to the 
central binary at almost all precessional phases 
except perhaps near $\Psi = 0$.   The high opacity of the disk 
in the edge-on configuration reduces the X-ray and optical 
eclipse visibility during that part of the precessional cycle. 
Therefore, the best opportunity to observe directly the companion star
is probably near $\Psi = 0$ when the disk obscuration 
attains a minimum (and not during the edge-on phases 
as suggested in the past; c.f. \citet{lei84}) and when the optical 
star is well in the foreground (near $\phi = 0$).  
We have a few spectra that satisfy these stringent timing requirements
(obtained over 3 nights beginning on HJD 2,451,463), 
and it is interesting to note that a central absorption 
feature was visible in the \ion{He}{1} features at that
time that migrated redward as expected for the orbital motion 
of the optical star (see Fig.~13).   The radial velocity shift 
is consistent with $K_O = 126 \pm 26$ km~s$^{-1}$, which, 
when combined with $K_X = 175 \pm 20$ km~s$^{-1}$ and $i=79$,  
yields $q=0.72 \pm 0.17$, $M_X/M_\odot = 16 \pm 6$, and $M_O/M_\odot = 23 \pm 8$.
However, other interpretations of this feature are possible, 
and additional radial velocities derived from photospheric absorption 
lines are needed to obtain reliable mass estimates. 
High resolution spectroscopy in the blue (where metallic 
absorption features are more common in B-type spectra) 
during times of favorable orientation might well yield 
a reliable detection of the optical companion. 

\placefigure{fig12}     


\acknowledgments

We thank the KPNO staff, and in particular Diane Harmer 
and Daryl Willmarth, for their assistance in making these 
observations with the KPNO Coude Feed Telescope, 
and we also thank Tammy Josephs for observing assistance at KPNO.  
We are grateful to the staff of McDonald Observatory and 
Tom Montemayor for their help at the 2.1-m telescope.   
This research has made use of results provided by the
ASM/RXTE teams at MIT and at the RXTE SOF and GOF at 
the NASA/Goddard Space Flight Center.
Support for this work was provided by NASA through grant number
GO-8308 from the Space Telescope Science Institute, which is
operated by the Association of Universities for Research in
Astronomy, Inc., under NASA contract NAS5-26555. 
Institutional support has been provided from the GSU College
of Arts and Sciences and from the Research Program Enhancement 
fund of the Board of Regents of the University System of Georgia,
administered through the GSU Office of the Vice President 
for Research and Sponsored Programs.  
We gratefully acknowledge all this support. 





\clearpage

\begin{figure}
\caption{The H$\alpha$ emission components from the jets over 
the last 7 nights of observation.  The left panel shows the 
approaching jet components and the right panel shows the 
receding components (with a reversed wavelength scale so that 
the Doppler shifts can be easily compared between panels). 
Each spectrum is placed so that the continuum in aligned with 
the time of observation and is scaled in intensity so that 
the continuum strength is equivalent to 0.5 d on the $y$-axis.
Tick marks indicate the central position of each component 
as determined by Gaussian fitting.} 
\label{fig1}
\end{figure}

\begin{figure}
\caption{The radial velocity curves of the H$\alpha -$ ({\it filled
circles}) and H$\alpha +$ ({\it open circles}) emission components
(symbols are plotted with areas proportional to $W_\lambda$). 
The global fit made using the \citet{kat82} nodding 
motions model is shown as a solid line.} 
\label{fig2}
\end{figure}

\begin{figure}
\caption{The ratio of the mean emission equivalent width for the 
approaching jet to that of the receding jet  for each of the 
observing runs plotted as a function of jet precessional phase.  
The continuous lines show the predicted amount of relativistic 
boosting for Doppler exponents 4 ({\it solid}) and 3 ({\it dotted}).} 
\label{fig3}
\end{figure}

\begin{figure}
\caption{Average H$\alpha$ stationary profiles plotted as 
a function of heliocentric radial velocity and date of run. 
The spectra are placed so that their continua are 
aligned with the mean date of observation, and the 
intensity scale in continuum units is given in the upper right
corner.  Each profile is labelled with the mean precessional 
phase.  The dashed line gives the profile from the 1998 run 
placed at the time corresponding to its precessional phase
in an earlier cycle.  The left hand panel illustrates the 
development of \ion{Na}{1} $\lambda\lambda 5890, 5896$
emission during the final two runs.} 
\label{fig4}
\end{figure}

\begin{figure}
\caption{Average \ion{He}{1} stationary profiles in the 
same format as Fig.~4.  The dashed line in the middle 
panel for \ion{He}{1} $\lambda 6678$ corresponds to the 
mean profile from the 1998 run plotted at a date 
corresponding to its precessional phase.} 
\label{fig5}
\end{figure}

\begin{figure}
\caption{Individual H$\alpha$ stationary profiles from our 
longest run plotted as a function of heliocentric wavelength 
and time of observation (in the same way as Fig.~4).
The corresponding orbital phase is indicated on the right hand
side.  The dashed line shows the fitted radial velocity curve for 
the H$\alpha +$ jet components (one of which appears near 6484\AA 
~at HJD 2,451,364; other, weaker jet components may be blended 
with the stationary H$\alpha$ line in the preceding two nights). 
The nearly vertical dotted line traces the outward motion of 
a blue-shifted absorption trough.} 
\label{fig6}
\end{figure}

\begin{figure}
\caption{Individual H$\alpha$ stationary profiles from our 
final run plotted in the same format as Fig.~6.
Here the dashed line shows the fitted radial velocity curve for 
the H$\alpha -$ jet, while the dashed - triple dotted line 
shows the same for jet features of \ion{He}{1} $\lambda 6678$. 
The dashed - single dotted line shows the radial velocity 
curve of the opposite red jet for \ion{He}{1} $\lambda 5876$. 
No \ion{He}{1} jet components are readily visible at their 
expected positions.  
The nearly vertical dotted line traces the outward motion of 
a red-shifted emission bump.} 
\label{fig7}
\end{figure}

\begin{figure}
\caption{Radial velocity curves of the stationary line emission peaks.
The measured relative velocities are shown for each observing run using 
the following symbols:
1 -- X, 
2 -- plus sign,
3 -- asterisk,
4 -- diamond,
5 -- triangle, and
6 -- square. 
The solid lines show the sinusoidal fits for each (Table 6).} 
\label{fig8}
\end{figure}

\begin{figure}
\caption{Radial velocity curve of the \ion{C}{2} $\lambda\lambda 7231, 7236$ 
stationary emission line.   Measurements from different runs 
are plotted using the same symbols as in Fig.~8.  
The solid line shows a sinusoidal fit to the entire set.} 
\label{fig9}
\end{figure}

\begin{figure}
\caption{The average spectrum of SS~433 ({\it top}) after 
removal of the H$\alpha$ jet components and division 
of the interstellar spectrum ({\it third from top}, 
offset in intensity by $-0.5$ for clarity). 
Also shown are the absorption spectrum from Earth's 
atmosphere ({\it bottom}, reduced by a factor of 2 in intensity 
and offset by $-1.0$) and the spectrum of 9~Cep (B2~Ib) 
({\it second from top}, offset in intensity by $-0.2$ for clarity).} 
\label{fig10}
\end{figure}

\begin{figure}
\caption{The observed and predicted spectral flux distribution 
of SS~433 in the ultraviolet and optical.  The arrows at the shorter 
wavelengths illustrate the flux upper limits from the 
STIS observation.  The line at the top of each arrow 
corresponds to the mean $+2\sigma_\mu$ flux limit over the 
wavelength range covered by the line.  The tip of the 
lower part of the arrow gives the measured mean flux in the band
(generally too small for reliable detection).   The two error
bars at 2770~\AA ~give the average bright and faint fluxes 
found by \citet{dol97} using {\it HST}/HSP.  The three error
bars between 2900~\AA ~and 3000~\AA ~show the mean fluxes in 
{\it IUE} spectrum, LWR4698.  The optical spectrophotometry 
from \citet{wag86} is averaged for the bright ({\it diamonds})
and faint ({\it squares}) states.  Model fluxes for reddened 
blackbody curves are shown for 
$T=$ 72,000, 45,000, 21,000~K and $A_V=8.4, 7.8, 7.8$ 
({\it dotted, dashed, solid lines}), respectively.} 
\label{fig11}
\end{figure}

\begin{figure}
\caption{
The RXTE/ASM X-ray count rates plotted against several 
of the SS~433 ``clocks''.   
{\it Top:} Binned averages as a function of precessional 
phase where $\Psi = 0$ corresponds to maximum disk opening 
(epoch of greatest velocity separation in the jet components).
{\it Second from top:} Data from near maximum disk opening 
plotted against orbital phase ($\phi=0$ is the epoch of the 
primary optical eclipse). 
{\it Third from top:} Data from the more edge-on precessional
phases plotted against orbital phase. 
{\it Bottom:} Data from near maximum disk opening 
plotted against nutational phase (0 corresponds to 
maximum optical light in the nodding cycle; \citet{gor98a}).
Error bars show the standard deviation of the mean in each case.} 
\label{fig12}
\end{figure}

\begin{figure}
\caption{\ion{He}{1} $\lambda 6678$ profiles obtained 
near $\Psi = 0$ and $\phi= 0$ plotted in a format similar to 
Fig.~6.  The tick marks indicate the positions of a 
central absorption feature that moved redward in this 
sequence in the manner expected for the optical star. 
The absorption feature disappeared on the subsequent 3 nights
when the optical star was behind the disk.} 
\label{fig13}
\end{figure}


\clearpage


\begin{deluxetable}{cccccl}
\tablewidth{0pc} 
\tablecaption{Journal of Optical Spectroscopy \label{tab1}}
\tablehead{
\colhead{Run} &
\colhead{Dates} &
\colhead{Range} &
\colhead{Resolution} &
\colhead{} &
\colhead{Observatory/Telescope/} \\ 
\colhead{Number} &
\colhead{(HJD-2,451,000)} &
\colhead{(\AA)} &
\colhead{($\lambda/\triangle\lambda$)} &
\colhead{Number} &
\colhead{Grating/CCD}} 
\startdata
1 &\phn53.7 -- \phn62.7& 6313 -- 6978       &\phn9530    &     5  &    KPNO/0.9m/B/F3KB     \\
2 &354.7 -- 354.9   &    6431 -- 6785       &\phn5440    &     2  &    KPNO/0.9m/RC181/TI5  \\   
2 &355.7 -- 364.9   &    5405 -- 6743       &\phn3950    &     9  &    KPNO/0.9m/RC181/F3KB \\
2 &355.9 -- 365.0   &    6461 -- 7799       &\phn4940    &     9  &    KPNO/0.9m/RC181/F3KB \\
3 &395.9 -- 399.9   &    5510 -- 6854       &  41960     &     3  &    McD/2.1m/Echelle/RA2 \\
4 &421.7 -- 429.7   &    5397 -- 6735       &\phn4050    &     7  &    KPNO/0.9m/RC181/F3KB \\
4 &421.7 -- 429.6   &    6453 -- 7791       &\phn4240    &     6  &    KPNO/0.9m/RC181/F3KB \\
5 &463.7 -- 467.7   &    5400 -- 6736       &\phn4100    &     5  &    KPNO/0.9m/RC181/F3KB \\
5 &463.6 -- 468.7   &    6446 -- 7782       &\phn5020    &     6  &    KPNO/0.9m/RC181/F3KB \\
6 &491.6 -- 497.6   &    5545 -- 6881       &\phn4400    &     7  &    KPNO/0.9m/RC181/F3KB \\
6 &491.6 -- 497.6   &    6590 -- 7927       &\phn5220    &     6  &    KPNO/0.9m/RC181/F3KB \\
\enddata
\end{deluxetable}

\clearpage


\begin{deluxetable}{crrc}
\tablewidth{0pc} 
\tablecaption{Jet Component Measurements \label{tab2}}
\tablehead{
\colhead{Date}     &
\colhead{$z$}               &
\colhead{$W_\lambda$} &
\colhead{FWHM} \\ 
\colhead{(HJD-2,451,000)}     &
\colhead{($\triangle\lambda/\lambda$)} &
\colhead{(\AA)} &
\colhead{(\AA)}} 
\startdata
\multispan{4}{\hfill H$\alpha -$ Components \hfill} \\
\vspace{-7pt} \\
\tableline
 \phn56.709 &   0.0503 &   -49 &    42 \\
 \phn57.732 &   0.0519 &   -44 &    43 \\
 \phn62.709 &   0.0634 &   -47 &    70 \\
 355.878 &   0.0499 &   -14 &    23 \\
 355.878 &   0.0560 &   -23 &    40 \\
 356.932 &   0.0588 &   -62 &    31 \\
 357.887 &   0.0602 &   -87 &    35 \\
 359.865 &   0.0538 &   -70 &    39 \\
 359.865 &   0.0587 &   -16 &    19 \\
 359.865 &   0.0626 &   -12 &    27 \\
 360.889 &   0.0526 &   -56 &    32 \\
 360.889 &   0.0580 &   -18 &    47 \\
 361.932 &   0.0530 &   -28 &    38 \\
 361.932 &   0.0640 &   -53 &    52 \\
 362.957 &   0.0529 &   -11 &    42 \\
 362.957 &   0.0656 &   -16 &    25 \\
 362.957 &   0.0720 &   -26 &    26 \\
 363.891 &   0.0649 &    -7 &    19 \\
 363.891 &   0.0718 &    -6 &    23 \\
 363.891 &   0.0787 &   -76 &    31 \\
 364.953 &   0.0599 &   -17 &    26 \\
 364.953 &   0.0655 &   -38 &    53 \\
 364.953 &   0.0781 &   -46 &    43 \\
 421.726 &  -0.0297 &   -19 &    61 \\
 424.717 &  -0.0303 &   -50 &    44 \\
 429.675 &  -0.0607 &   -24 &    36 \\
 463.674 &  -0.0901 &   -67 &    31 \\
 464.669 &  -0.0975 &   -19 &    30 \\
 464.669 &  -0.0935 &   -26 &    26 \\
 464.669 &  -0.0896 &   -11 &    22 \\
 465.674 &  -0.1039 &   -37 &    29 \\
 465.674 &  -0.0978 &   -12 &    17 \\
 465.674 &  -0.0923 &   -20 &    57 \\
 466.660 &  -0.1055 &   -42 &    35 \\
 466.660 &  -0.0974 &    -4 &    16 \\
 466.660 &  -0.0925 &   -10 &    53 \\
 467.672 &  -0.1058 &   -31 &    33 \\
 467.672 &  -0.0964 &   -35 &    74 \\
 491.644 &  -0.0500 &   -43 &    40 \\
 492.622 &  -0.0497 &   -13 &    42 \\
 492.622 &  -0.0333 &   -13 &    24 \\
 492.622 &  -0.0259 &   -19 &    30 \\
 493.612 &  -0.0218 &   -40 &    26 \\
 494.620 &  -0.0225 &   -83 &    31 \\
 495.623 &  -0.0249 &   -77 &    32 \\
 496.619 &  -0.0258 &   -60 &    26 \\
 497.623 &  -0.0266 &   -57 &    29 \\
\tableline
\vspace{-7pt} \\
\multispan{4}{\hfill H$\alpha +$ Components \hfill} \\
\vspace{-7pt} \\
\tableline
 \phn53.674 &  -0.0095 &   -50 &    63 \\
 \phn55.705 &  -0.0118 &   -51 &    45 \\
 \phn56.709 &  -0.0124 &    -6 &    40 \\
 \phn62.709 &   0.0148 &    -5 &    28 \\
 354.740 &   0.0266 &   -45 &    32 \\
 354.925 &   0.0259 &   -73 &    41 \\
 354.740 &   0.0182 &   -40 &    21 \\
 355.878 &   0.0181 &   -48 &    21 \\
 355.878 &   0.0266 &   -18 &    36 \\
 356.813 &   0.0121 &   -14 &    17 \\
 356.932 &   0.0118 &   -17 &    17 \\
 357.769 &   0.0116 &   -29 &    24 \\
 357.887 &   0.0119 &   -35 &    25 \\
 359.747 &   0.0192 &   -12 &    29 \\
 359.865 &   0.0268 &   -53 &    42 \\
 359.865 &   0.0193 &   -40 &    29 \\
 360.697 &   0.0155 &   -48 &    43 \\
 360.889 &   0.0271 &   -10 &    33 \\
 360.889 &   0.0171 &   -13 &    28 \\
 360.889 &   0.0131 &    -8 &    22 \\
 361.895 &   0.0135 &    -2 &    28 \\
 361.932 &   0.0132 &    -1 &    11 \\
 363.716 &  -0.0124 &   -50 &    25 \\
 363.891 &  -0.0120 &   -40 &    22 \\
 364.921 &  -0.0115 &   -14 &    27 \\
 364.953 &  -0.0115 &    -8 &    22 \\
 399.890 &   0.0112 &   -36 &    24 \\
 421.670 &   0.1274 &    -7 &    38 \\
 421.670 &   0.0901 &    -5 &    28 \\
 421.670 &   0.0841 &    -5 &    29 \\
 424.642 &   0.1078 &   -41 &    42 \\
 429.624 &   0.1423 &   -15 &    50 \\
 463.619 &   0.1533 &   -57 &    50 \\
 464.618 &   0.1600 &   -30 &    41 \\
 464.618 &   0.1529 &   -14 &    32 \\
 465.617 &   0.1686 &   -26 &    75 \\
 465.617 &   0.1601 &   -14 &    43 \\
 465.617 &   0.1525 &    -6 &    28 \\
 466.611 &   0.1737 &   -21 &    55 \\
 466.611 &   0.1602 &   -10 &    49 \\
 466.611 &   0.1518 &    -3 &    29 \\
 467.612 &   0.1738 &   -26 &    65 \\
 467.612 &   0.1641 &   -19 &    91 \\
 468.687 &   0.1731 &   -10 &    54 \\
 468.687 &   0.1585 &    -7 &    32 \\
 468.687 &   0.1521 &   -33 &    38 \\
 491.599 &   0.1257 &   -26 &    33 \\
 492.571 &   0.1251 &    -8 &    27 \\
 492.571 &   0.1095 &   -13 &    36 \\
 492.571 &   0.1018 &   -32 &    39 \\
 494.570 &   0.0963 &   -65 &    33 \\
 495.570 &   0.0987 &   -76 &    59 \\
 496.571 &   0.0976 &   -71 &    48 \\
 497.571 &   0.0998 &   -51 &    49 \\
\enddata
\end{deluxetable}

\clearpage


\begin{deluxetable}{lccc}
\tablewidth{0pc} 
\tablecaption{Nodding and Precessional Model Parameters \label{tab3}}
\tablehead{
\colhead{Parameter} &
\colhead{H$\alpha -$} &
\colhead{H$\alpha +$} &
\colhead{Adopted}} 
\startdata
$N$                          &  47    &  54    & \nodata          \\
$t_{\rm or}$ (HJD-2,451,000) & 460.44 & 460.39 & $460.42\pm 0.04$ \\
$\xi_0$ ($^\circ$)           &  84.6  &  85.2  & $84.9\pm 0.5$    \\
$\triangle \xi$ ($^\circ$)   &  5.3   &  5.8   & $5.5\pm0.5$      \\
r.m.s. ($z$)                 & 0.008  & 0.011  & \nodata          \\
\enddata
\end{deluxetable}

\clearpage


\begin{deluxetable}{crrrr}
\tablewidth{0pc} 
\tablecaption{H$\alpha$ Stationary Emission Fits \label{tab4}}
\tablehead{
\colhead{Date} &
\colhead{$W_\lambda$} &
\colhead{FWHM} &
\colhead{$V_{\rm centroid}$} &
\colhead{$V_{\rm peak}$} \\
\colhead{(HJD-2,451,000)} &
\colhead{(\AA )} &
\colhead{(\AA )} &
\colhead{(km s$^{-1}$)} &
\colhead{(km s$^{-1}$)} 
} 
\startdata
 \phn53.674 &  -342 &  16.0 &   204 &   233 \\
 \phn55.677 &  -312 &  14.2 &   302 &   276 \\
 \phn56.709 &  -376 &  14.7 &   255 &   280 \\
 \phn57.721 &  -388 &  14.8 &   133 &   251 \\
 \phn62.699 &   -88 &  15.1 &   291 &   173 \\
 354.740 &  -200 &  20.8 &   365 &   273 \\
 354.925 &  -282 &  21.1 &   367 &   277 \\
 355.738 &  -296 &  17.6 &   420 &   314 \\
 355.878 &  -299 &  17.4 &   383 &   321 \\
 356.813 &  -306 &  17.3 &   480 &   337 \\
 356.932 &  -301 &  16.5 &   483 &   341 \\
 357.769 &  -317 &  17.4 &   475 &   355 \\
 357.887 &  -306 &  17.1 &   449 &   370 \\
 359.747 &  -303 &  19.4 &   515 &   328 \\
 359.865 &  -282 &  18.8 &   488 &   345 \\
 360.697 &  -305 &  22.7 &   519 &   313 \\
 360.889 &  -298 &  23.7 &   493 &   313 \\
 361.895 &  -337 &  24.9 &   415 &   318 \\
 361.932 &  -324 &  25.3 &   384 &   314 \\
 363.716 &  -295 &  25.2 &   234 &   166 \\
 363.891 &  -260 &  24.2 &   282 &   196 \\
 364.921 &  -378 &  22.5 &   718 &   151 \\
 364.953 &  -245 &  22.4 &   533 &   172 \\
 395.869 &  -265 &  15.8 &   445 &   254 \\
 398.932 &  -216 &  13.5 &   323 &   229 \\
 399.890 &  -246 &  14.8 &   387 &   249 \\
 421.670 &  -105 &  12.4 &   141 &   182 \\
 421.726 &  -108 &  12.4 &   136 &   179 \\
 423.722 &  -213 &  13.1 &   237 &   182 \\
 424.642 &  -227 &  17.7 &   258 &   168 \\
 424.717 &  -286 &  17.6 &   343 &   179 \\
 425.688 &  -204 &  14.4 &   244 &   185 \\
 425.743 &  -223 &  13.8 &   284 &   177 \\
 426.632 &  -157 &  13.5 &   234 &   173 \\
 427.704 &  -177 &  13.2 &   246 &   146 \\
 428.617 &  -157 &  12.2 &    -4 &    57 \\
 428.677 &  -193 &  12.4 &   171 &   145 \\
 429.624 &  -171 &  13.4 &    98 &   102 \\
 429.675 &  -216 &  13.6 &   195 &   139 \\
 463.619 &  -336 &  15.1 &   163 &   168 \\
 463.674 &  -338 &  15.1 &   176 &   176 \\
 464.618 &  -246 &  15.3 &   158 &   147 \\
 464.669 &  -240 &  15.2 &   178 &   160 \\
 465.617 &  -226 &  15.5 &   119 &   119 \\
 465.674 &  -211 &  15.5 &   117 &   117 \\
 466.611 &  -208 &  16.1 &    24 &    69 \\
 466.660 &  -199 &  16.0 &    43 &    76 \\
 467.612 &  -205 &  15.1 &    49 &    65 \\
 467.672 &  -195 &  14.8 &    51 &    65 \\
 468.687 &  -202 &  14.6 &    74 &    60 \\
 491.644 &  -368 &  14.6 &   335 &   150 \\
 492.622 &  -377 &  15.1 &   126 &   129 \\
 493.612 &  -376 &  15.5 &   158 &   119 \\
 494.621 &  -379 &  16.7 &   159 &   111 \\
 495.623 &  -363 &  21.7 &   127 &    70 \\
 496.619 &  -329 &  26.5 &   -10 &    11 \\
 497.623 &  -255 &  26.0 &    28 &    52 
\enddata
\end{deluxetable}

\clearpage


\begin{deluxetable}{crrrr}
\tablewidth{0pc} 
\tablecaption{\ion{He}{1} Stationary Emission Fits \label{tab5}}
\tablehead{
\colhead{Date} &
\colhead{$W_\lambda$} &
\colhead{FWHM} &
\colhead{$V_{\rm centroid}$} &
\colhead{$V_{\rm peak}$} \\
\colhead{(HJD-2,451,000)} &
\colhead{(\AA )} &
\colhead{(\AA )} &
\colhead{(km s$^{-1}$)} &
\colhead{(km s$^{-1}$)} 
} 
\startdata
\multispan{5}{\hfill \ion{He}{1} $\lambda 5876$ \hfill} \\
\vspace{-7pt} \\
\tableline
 355.738 &   -3.2 &    4.8 &   151 &   193 \\
 356.813 &   -4.6 &   11.0 &    65 &   205 \\
 359.747 &   -5.3 &    8.5 &   240 &   274 \\
 360.697 &   -4.6 &    9.2 &   157 &   219 \\
 362.839 &   -4.2 &    3.5 &    34 &   114 \\
 363.716 &   -2.9 &    5.0 &    53 &    71 \\
 364.921 &   -5.2 &    6.2 &     7 &    68 \\
 395.869 &   -8.8 &    8.4 &   259 &   263 \\
 399.890 &   -4.0 &    6.2 &   227 &   248 \\
 421.726 &   -4.1 &    6.2 &   139 &   209 \\
 423.722 &  -10.2 &    6.8 &    93 &   186 \\
 424.717 &   -8.9 &   10.0 &   104 &   194 \\
 425.743 &   -7.1 &    8.4 &   107 &   177 \\
 427.704 &   -6.9 &    9.3 &    22 &   125 \\
 428.677 &   -7.5 &    8.5 &    16 &   115 \\
 429.675 &  -11.5 &    9.1 &    80 &   164 \\
 463.674 &  -11.9 &    9.5 &   250 &   206 \\
 464.669 &   -7.5 &   10.6 &   207 &   183 \\
 491.644 &  -12.2 &   13.1 &   175 &   152 \\
 494.621 &   -7.7 &   12.1 &   196 &   145 \\
 495.623 &   -7.6 &    7.1 &   101 &    90 \\
 496.619 &   -7.5 &    7.7 &   100 &   111 \\
 497.623 &   -5.2 &    8.0 &   205 &   178 \\
\tableline
\vspace{-7pt} \\
\multispan{5}{\hfill \ion{He}{1} $\lambda 6678$ \hfill} \\
\vspace{-7pt} \\
\tableline
 \phn53.674 &   -8.3 &    6.7 &   303 &   226 \\
 \phn55.677 &   -8.2 &    8.0 &   251 &   237 \\
 \phn56.709 &  -10.1 &    9.8 &   307 &   241 \\
 \phn57.721 &  -10.4 &   10.3 &   316 &   234 \\
 354.740 &   -1.7 &    6.0 &   210 &   219 \\
 362.839 &   -8.4 &   18.5 &    66 &    -3 \\
 363.716 &   -4.9 &    9.4 &    40 &    85 \\
 363.891 &   -3.2 &    8.0 &    85 &    91 \\
 364.921 &   -6.4 &    9.0 &    96 &    86 \\
 364.953 &   -4.7 &    8.4 &   130 &   112 \\
 421.670 &   -2.7 &    4.2 &    45 &   199 \\
 424.642 &   -5.6 &    8.6 &   -52 &   192 \\
 424.717 &   -1.8 &    4.8 &   123 &   199 \\
 425.688 &   -4.3 &    5.7 &    53 &   162 \\
 425.743 &   -2.5 &    4.8 &   139 &   178 \\
 426.632 &   -3.3 &    9.7 &    77 &   180 \\
 427.704 &   -2.8 &   10.0 &    14 &   116 \\
 428.617 &   -4.3 &   10.3 &   -79 &    96 \\
 428.677 &   -4.8 &   10.8 &    -6 &   133 \\
 429.624 &   -7.5 &    6.7 &   -99 &   126 \\
 429.675 &   -4.8 &    6.6 &    -6 &   137 \\
 463.674 &   -9.6 &    9.1 &   178 &   172 \\
 464.618 &   -7.1 &   10.9 &    81 &    89 \\
 464.669 &   -7.0 &   11.1 &    91 &    86 \\
 465.617 &   -7.4 &   11.8 &   -24 &    17 \\
 465.674 &   -6.9 &   11.6 &   -31 &    13 \\
 466.611 &   -6.7 &    9.9 &   -82 &   -32 \\
 466.660 &   -6.5 &    9.7 &   -82 &   -37 \\
 467.612 &   -5.9 &    9.4 &   -41 &   -22 \\
 467.672 &   -5.7 &    9.3 &   -54 &   -18 \\
 468.687 &   -5.3 &   10.0 &   -73 &   -28 \\
 491.599 &   -6.0 &   11.4 &    99 &   119 \\
 493.612 &   -5.2 &   11.3 &    66 &    44 \\
 494.570 &   -4.2 &   11.0 &   135 &    81 \\
 494.621 &   -4.1 &    8.2 &   108 &    66 \\
 495.570 &   -5.5 &    9.4 &    60 &    59 \\
 495.623 &   -5.2 &    8.5 &    74 &    66 \\
 496.571 &   -5.6 &    8.9 &    70 &    59 \\
 496.619 &   -5.2 &    9.3 &    41 &    62 \\
 497.571 &   -4.7 &    8.4 &    54 &    68 \\
 497.623 &   -4.9 &    8.5 &    89 &    83 \\
\tableline
\vspace{-7pt} \\
\multispan{5}{\hfill \ion{He}{1} $\lambda 7065$ \hfill} \\
\vspace{-7pt} \\
\tableline
 355.878 &  -16.2 &   11.0 &   464 &   301 \\
 356.932 &  -10.5 &   11.0 &   368 &   328 \\
 357.887 &  -12.0 &   12.9 &   217 &   319 \\
 359.865 &  -12.8 &   11.2 &   531 &   291 \\
 360.889 &  -10.6 &   12.6 &   487 &   246 \\
 361.932 &  -10.0 &   14.2 &   138 &   225 \\
 421.670 &   -2.7 &    5.8 &   114 &   206 \\
 424.642 &  -14.1 &   10.6 &  -128 &   186 \\
 425.688 &   -7.7 &    8.6 &   -49 &   177 \\
 426.632 &   -6.1 &   11.0 &   -56 &   158 \\
 428.617 &   -7.2 &   10.3 &  -130 &    95 \\
 429.624 &   -9.9 &    8.2 &   -92 &   124 \\
 463.619 &  -16.4 &   11.5 &   124 &   176 \\
 464.618 &  -10.1 &   12.5 &   151 &   158 \\
 465.617 &   -9.8 &   13.0 &   100 &   114 \\
 466.611 &   -9.3 &   13.5 &    61 &    49 \\
 467.612 &   -8.5 &   12.6 &    62 &    36 \\
 468.687 &   -9.2 &   12.7 &    -5 &     3 \\
 494.570 &   -6.7 &   13.5 &   120 &    79 \\
 495.570 &   -9.3 &   12.2 &   253 &    90 \\
 496.571 &   -9.9 &   10.6 &   188 &    79 \\
 497.571 &   -7.5 &    9.8 &   167 &   117 
\enddata
\end{deluxetable}

\clearpage


\begin{deluxetable}{lcccc}
\tablewidth{0pc} 
\tablecaption{\ion{He}{1} Stationary Emission Peak Radial Velocity Fits \label{tab6}}
\tablehead{
\colhead{Element} &
\colhead{H$\alpha$} &
\colhead{\ion{He}{1} $\lambda 5876$} &
\colhead{\ion{He}{1} $\lambda 6678$} &
\colhead{\ion{He}{1} $\lambda 7065$}} 
\startdata
$K$ (km s$^{-1}$)    &    67 (5)  &       59 (6)        &         60 (5)      &         71 (5)      \\
$\phi$(max.)         &  0.04 (3)  &     0.01 (5)        &    $-0.03$ (3)      &    $-0.01$ (3)      \\
$V_1$ (km s$^{-1}$)  &   219      &   \nodata           &        194          &     \nodata         \\
$V_2$ (km s$^{-1}$)  &   281      &      167            &        143          &        252          \\
$V_3$ (km s$^{-1}$)  &   206      &      225            &     \nodata         &     \nodata         \\
$V_4$ (km s$^{-1}$)  &   152      &      163            &        163          &        157          \\
$V_5$ (km s$^{-1}$)  &   113      &      154            &         31          &         97          \\
$V_6$ (km s$^{-1}$)  &   130      &      170            &        112          &        148          \\
r.m.s. (km s$^{-1}$) &    27      &       31            &         21          &         16          \\
\enddata
\end{deluxetable}

\clearpage


\begin{deluxetable}{lccccc}
\tablewidth{0pc} 
\tablecaption{UV Flux Limits ($10^{-18}$ erg cm$^{-2}$ s$^{-1}$ \AA $^{-1}$) \label{tab7}}
\tablehead{
\colhead{Range (\AA )} &
\colhead{$<\lambda >$ (\AA )} &
\colhead{$n$} &
\colhead{$F_\lambda$} &
\colhead{$\sigma _\mu$} &
\colhead{$F_\lambda + 2 \sigma _\mu$} } 
\startdata
1155 -- 1185 &    1170.08 & 52 & \phs0.8 & \phn3.9  & \phn8.6  \\
1195 -- 1235 &    1215.05 & 68 & \phs3.5 &    14.8  &    33.0  \\
1250 -- 1275 &    1262.65 & 43 &  $-1.3$ & \phn1.2  & \phn2.5  \\
1285 -- 1320 &    1302.65 & 60 & \phs1.6 & \phn1.6  & \phn4.8  \\
1340 -- 1380 &    1359.88 & 68 & \phs3.3 & \phn1.4  & \phn6.1  \\
1400 -- 1450 &    1425.00 & 85 & \phs0.8 & \phn0.6  & \phn2.1  \\
1450 -- 1500 &    1474.93 & 86 & \phs0.6 & \phn0.8  & \phn2.2  \\
1500 -- 1550 &    1525.16 & 86 & \phs1.1 & \phn0.9  & \phn2.8  \\
1550 -- 1600 &    1575.09 & 85 & \phs1.8 & \phn1.2  & \phn4.2  
\enddata
\end{deluxetable}



\clearpage

\setcounter{figure}{0}
\begin{figure}[t]
\plotone{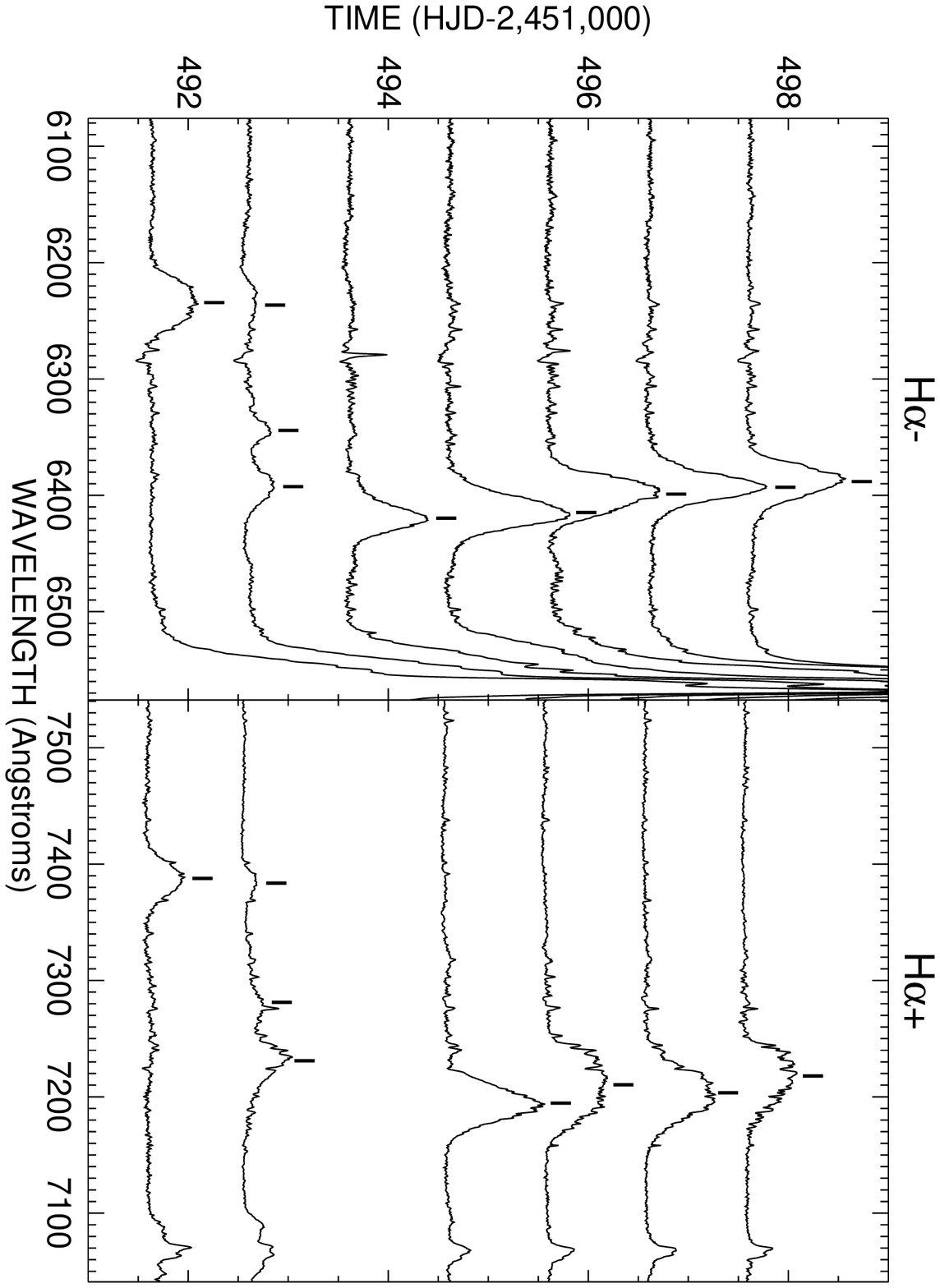}
\caption{}
\end{figure}

\begin{figure}[t]
\plotone{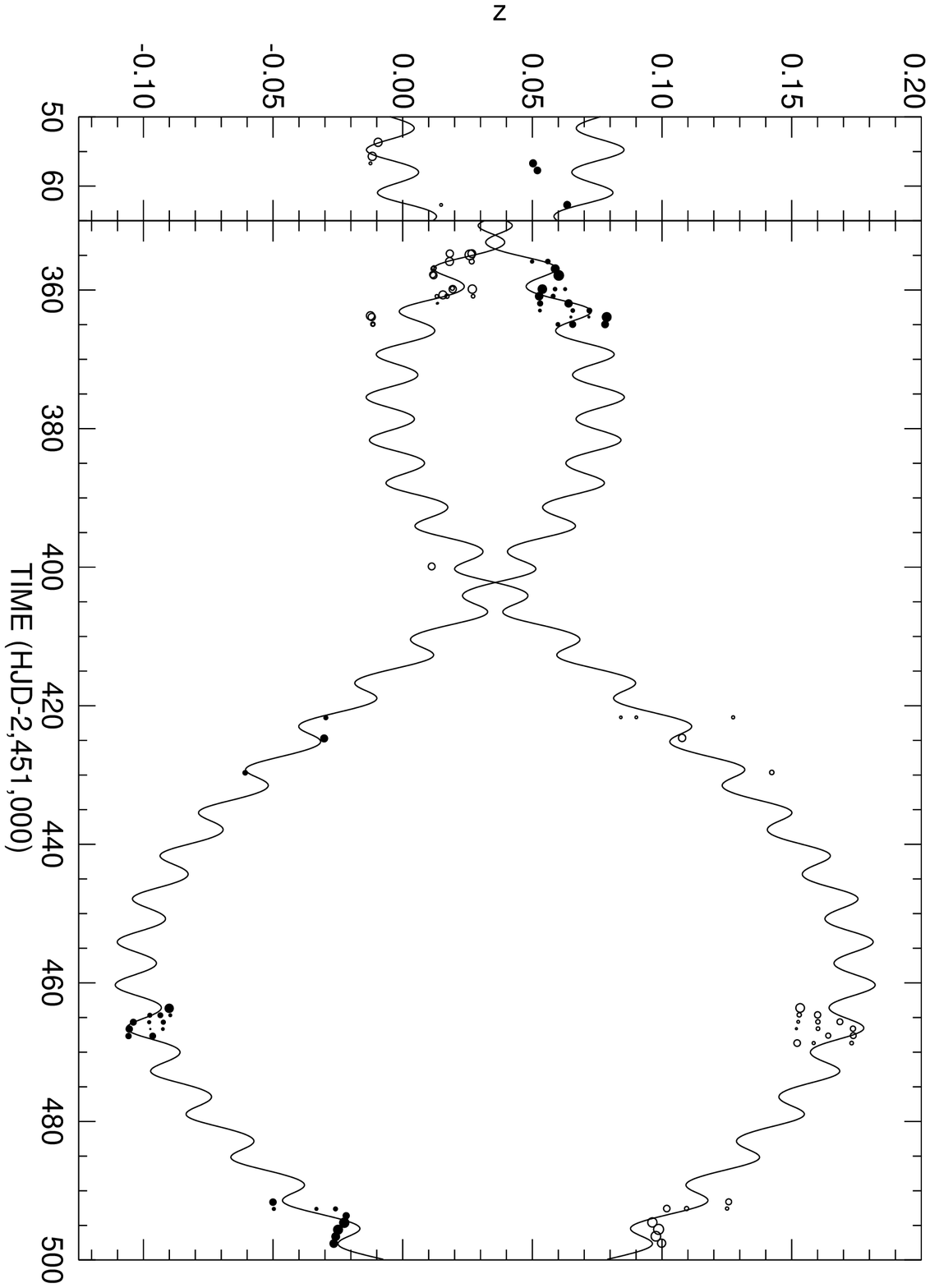}
\caption{}
\end{figure}

\begin{figure}[t]
\plotone{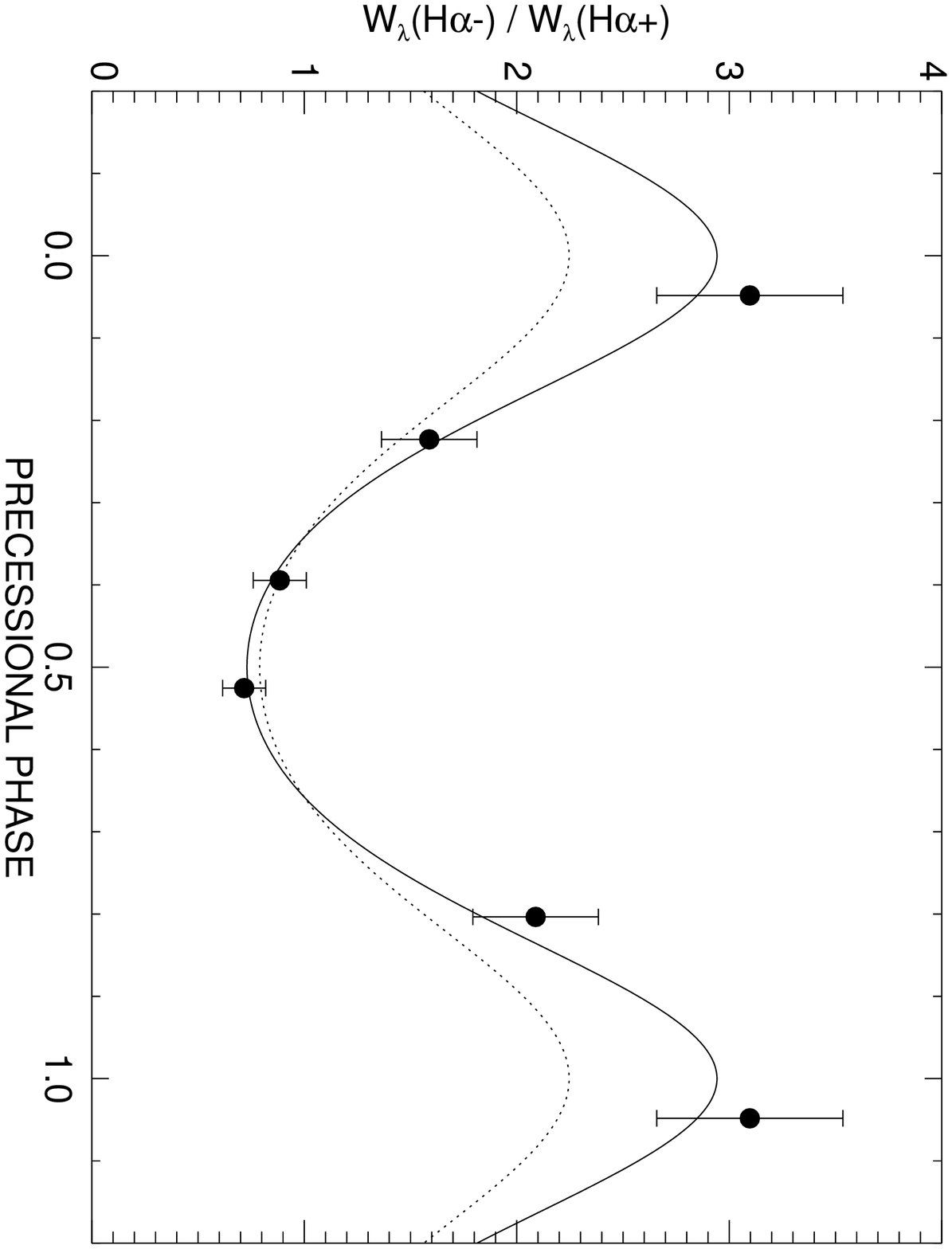}
\caption{}
\end{figure}

\begin{figure}[t]
\plotone{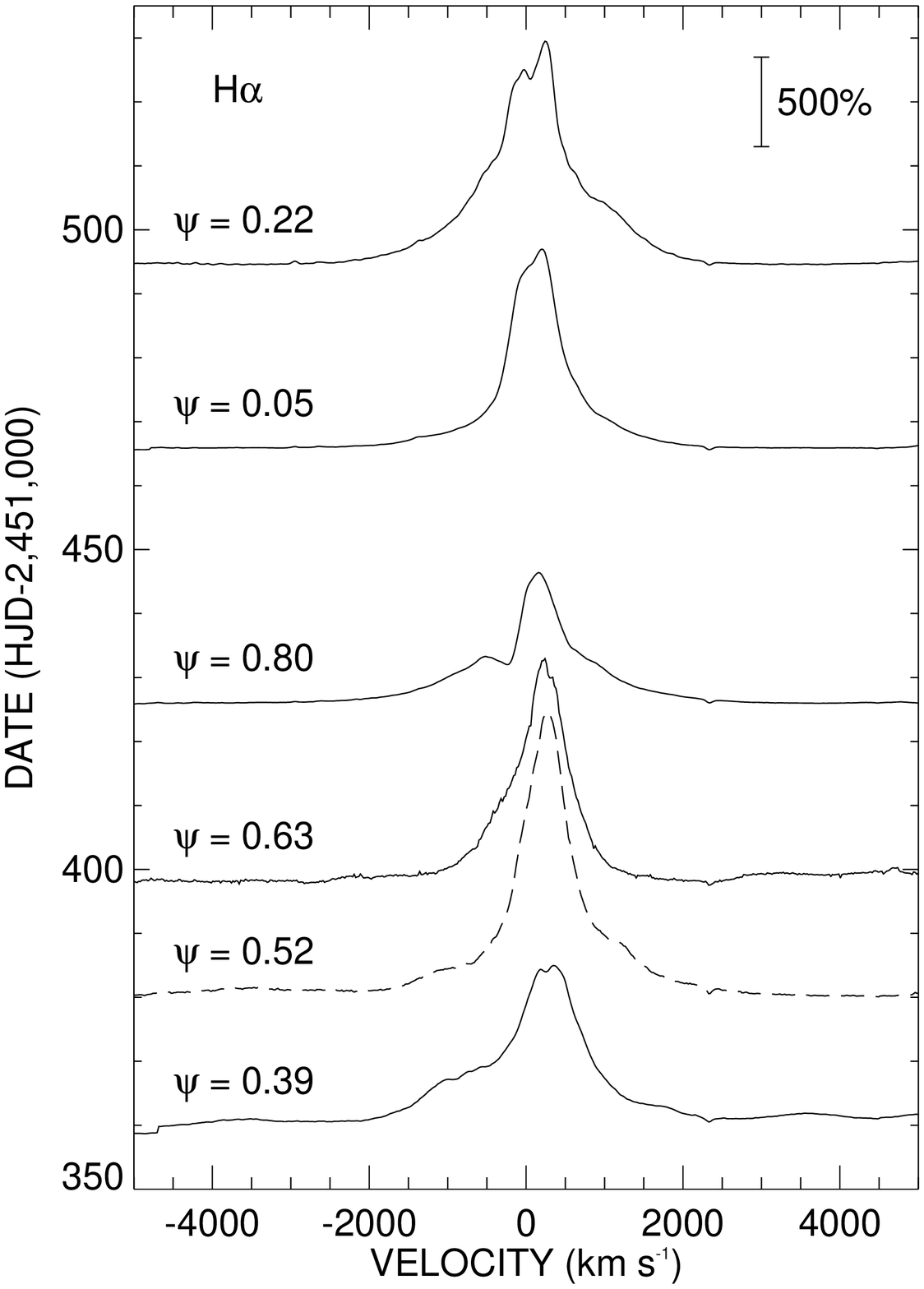}
\caption{}
\end{figure}

\begin{figure}[t]
\plotone{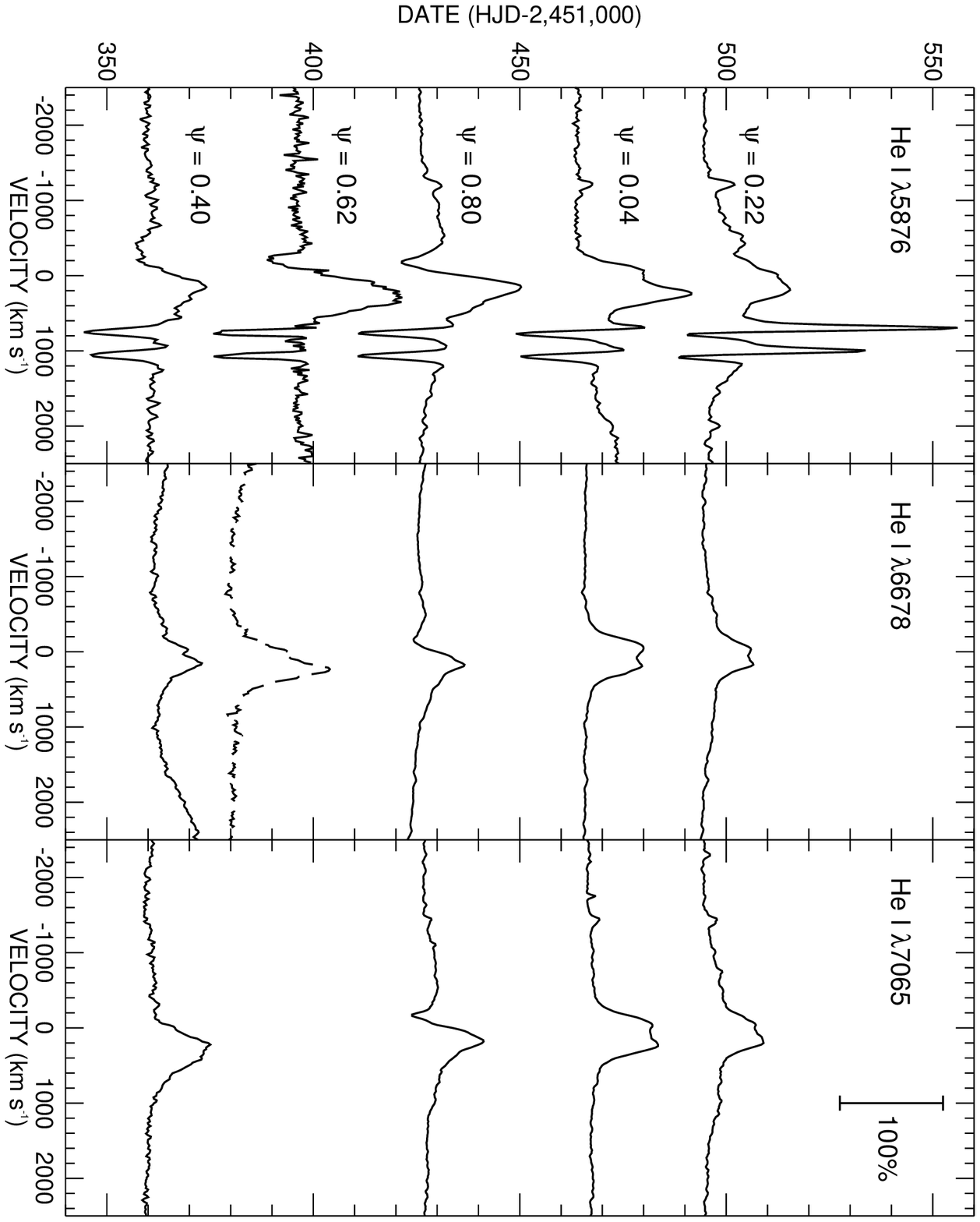}
\caption{}
\end{figure}

\begin{figure}[t]
\plotone{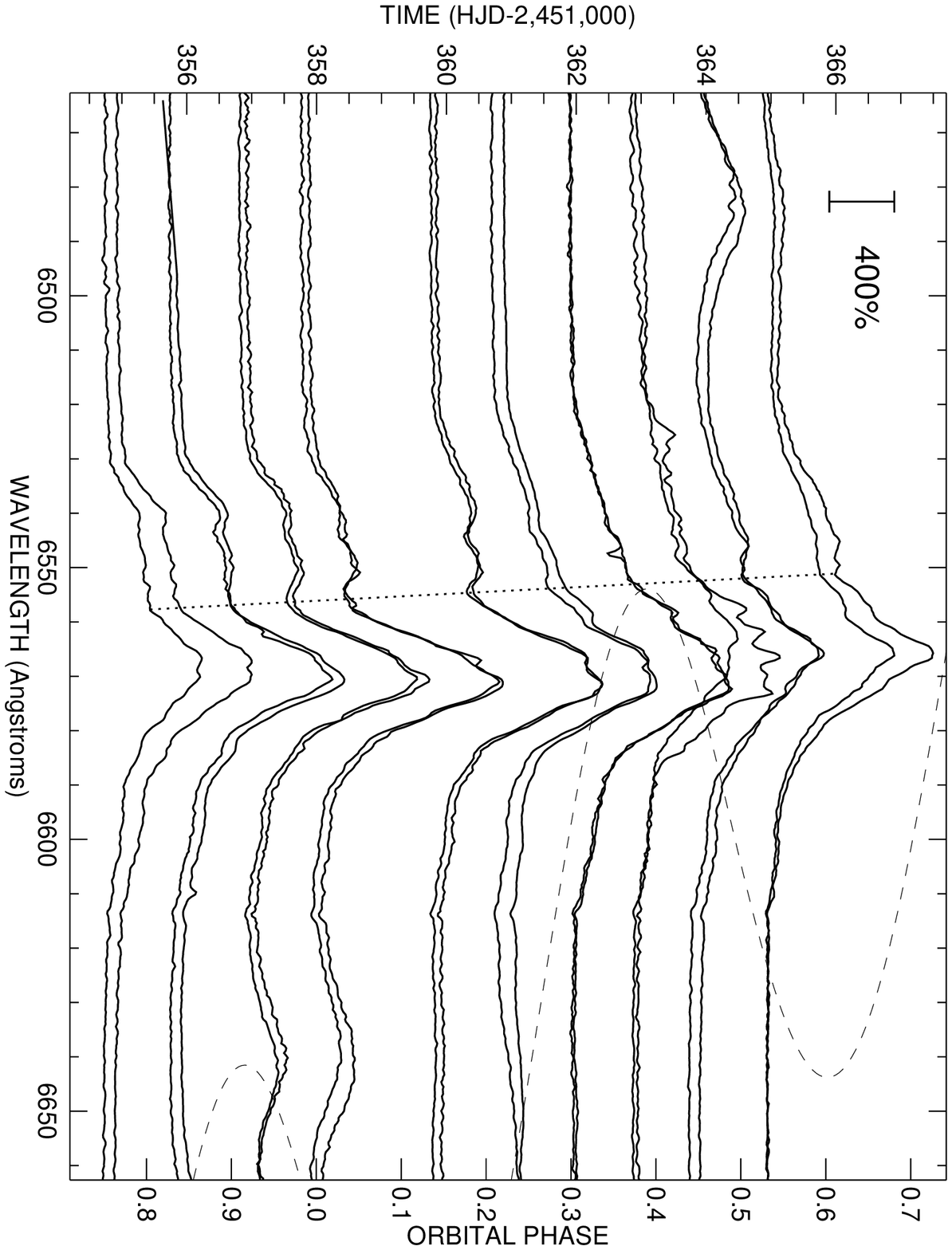}
\caption{}
\end{figure}

\begin{figure}[t]
\plotone{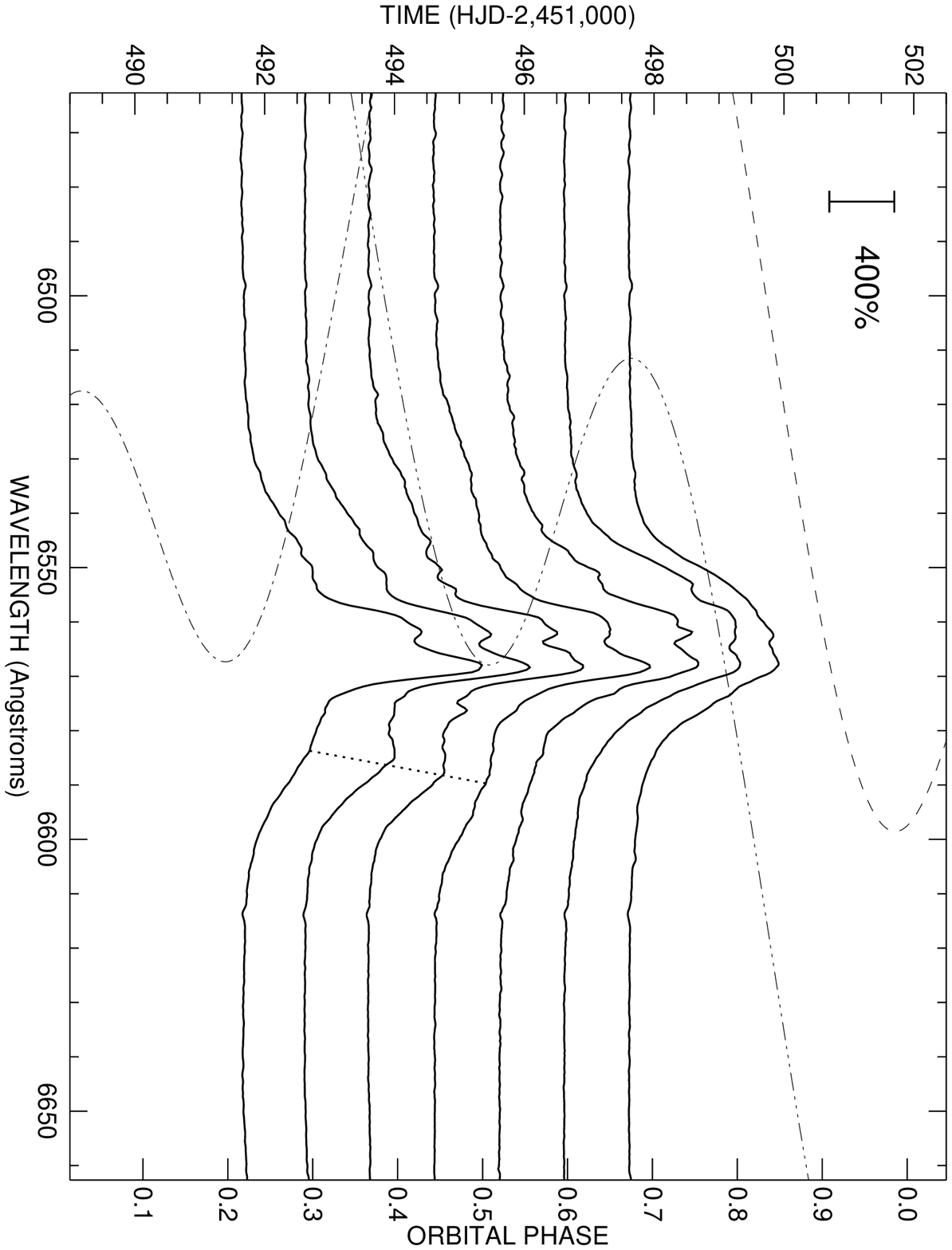}
\caption{}
\end{figure}

\begin{figure}[t]
\plotone{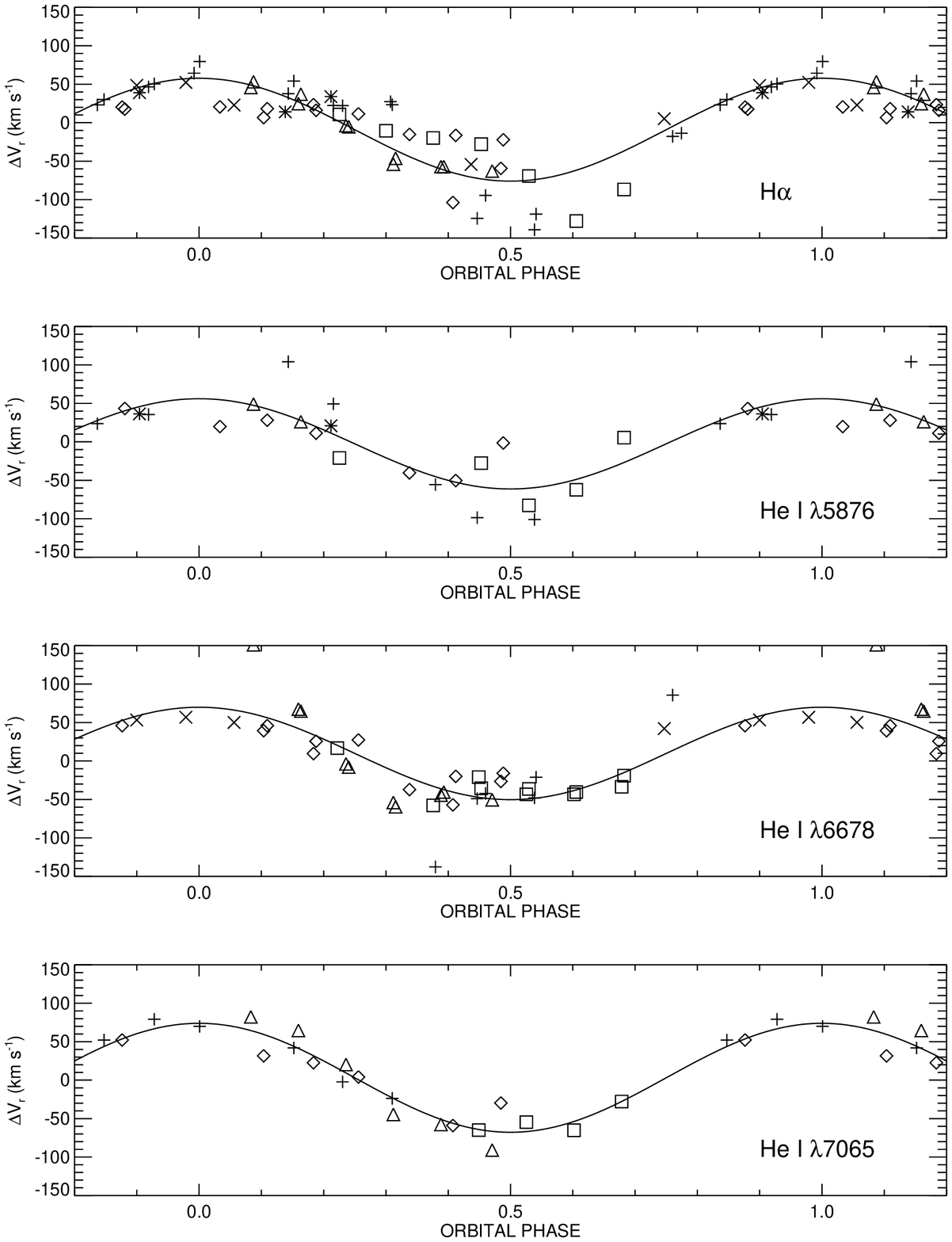}
\caption{}
\end{figure}

\begin{figure}[t]
\plotone{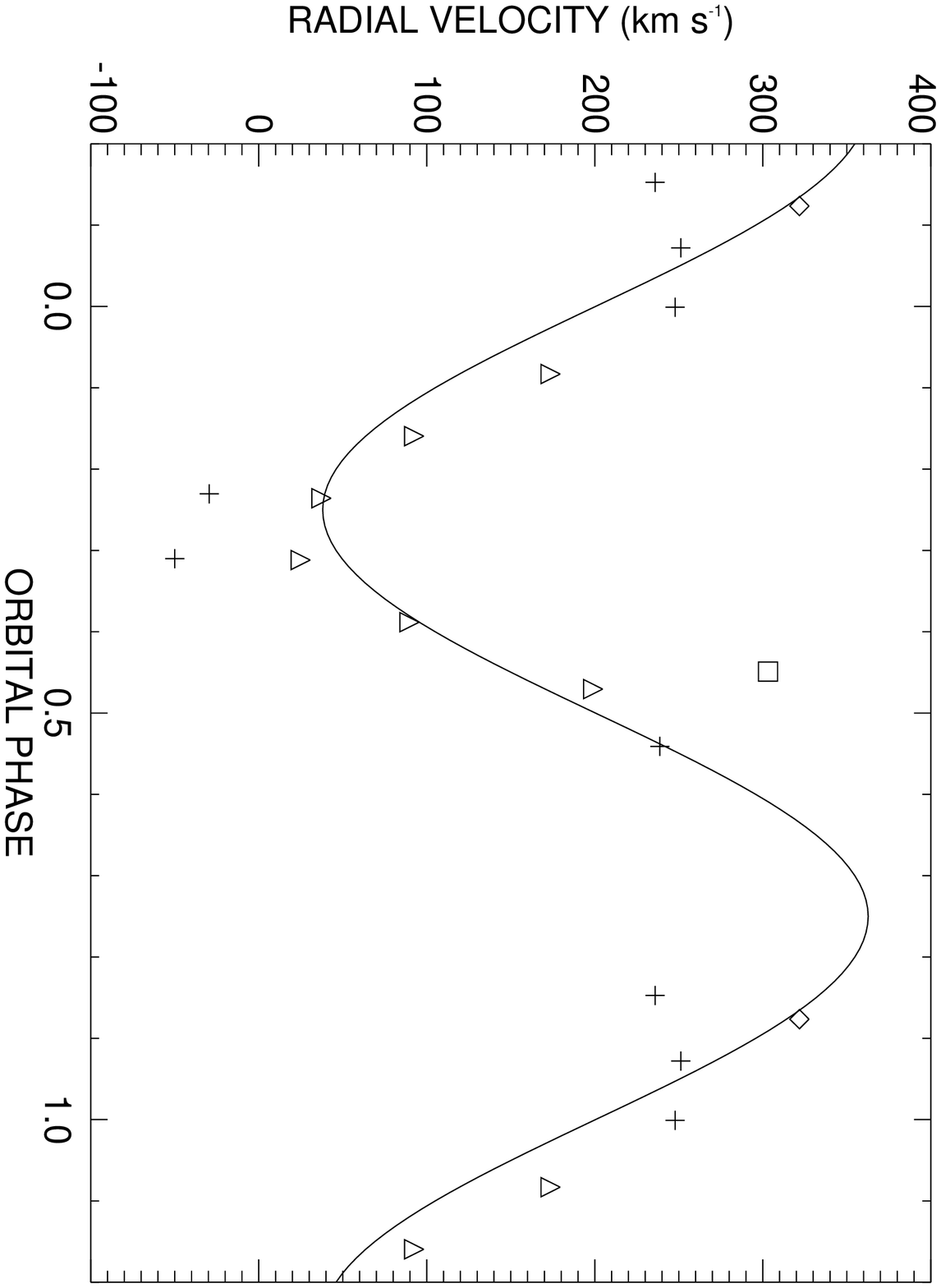}
\caption{}
\end{figure}

\begin{figure}[t]
\plotone{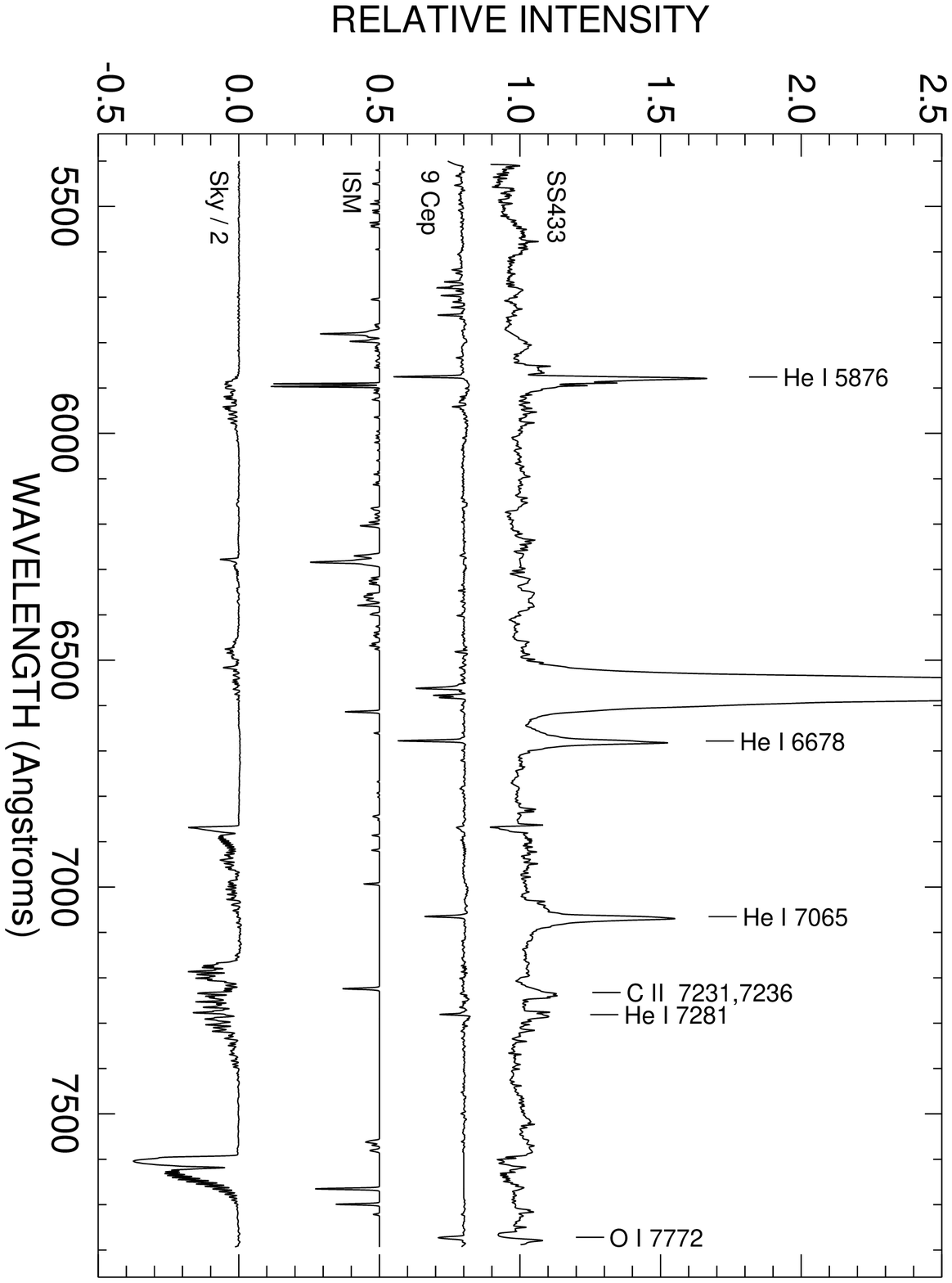}
\caption{}
\end{figure}

\begin{figure}[t]
\plotone{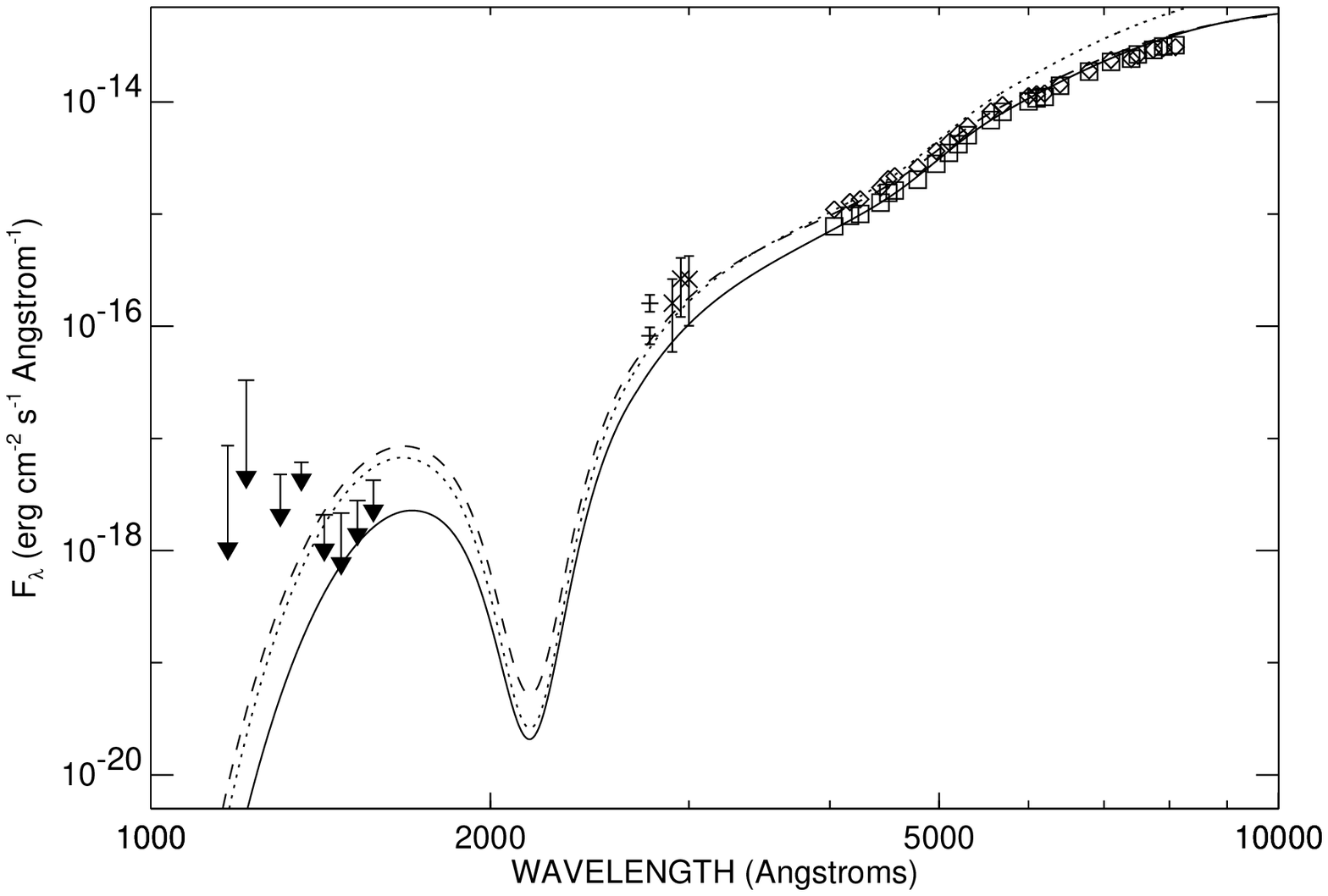}
\caption{}
\end{figure}

\begin{figure}[t]
\plotone{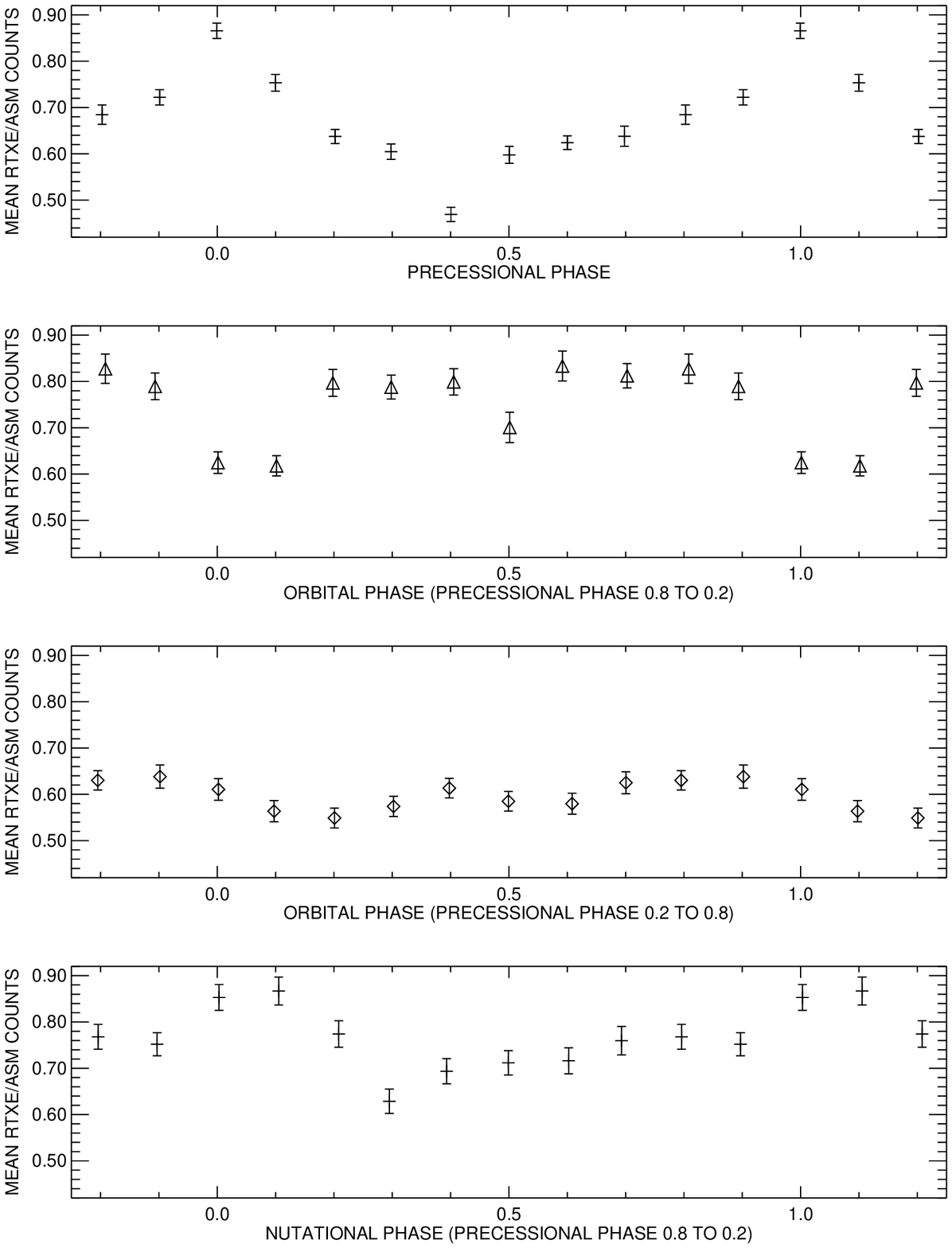}
\caption{}
\end{figure}

\begin{figure}[t]
\plotone{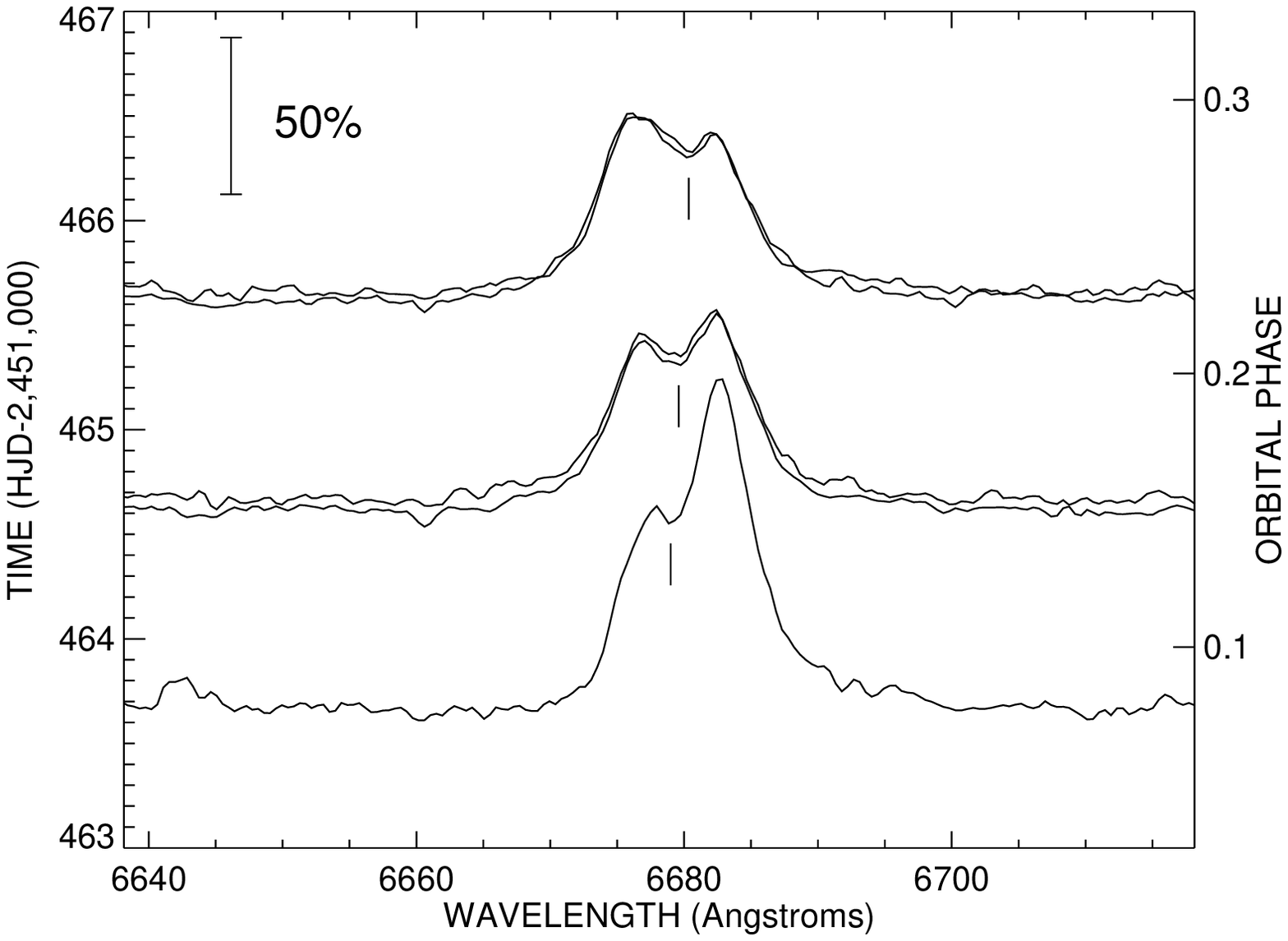}
\caption{}
\end{figure}


\end{document}